\documentclass{article}

\usepackage[english]{babel}

\usepackage[
  letterpaper,
  top=2cm,
  bottom=2cm,
  left=3cm,
  right=3cm,
  marginparwidth=1.75cm
]{geometry}

\usepackage{amsmath}
\usepackage{amssymb}
\usepackage{graphicx}
\usepackage{subcaption}
\usepackage{float}
\usepackage[colorlinks=true, allcolors=blue]{hyperref}

\title{When Certainty Is Not Worth It: Capital Lock-Up and Settlement Discounting in Prediction Markets}
\author{
  Jonas Gebele\\
  Technical University of Munich\\
  \texttt{jonas.gebele@tum.de}
  \and
  Florian Matthes\\
  Technical University of Munich\\
  \texttt{matthes@tum.de}
}

\begin{document}
  \maketitle

  \begin{abstract}
    Collateralized prediction markets are contingent-claim markets in which
    economic uncertainty can disappear before winning claims become redeemable.
    This paper studies the pricing effect of that delay. When collateral remains
    locked until oracle settlement, a near-certain dollar is a delayed dollar,
    so prices embed a maturity-dependent settlement discount in addition to
    beliefs about outcomes. We recover an implied settlement-discount term
    structure from persistent near-certain contracts using realized settlement
    times and summarize it as an annualized settlement wedge (ASW). The recovered
    wedges are positive, maturity-dependent, and time-varying. Adjusting prices
    by these curves reduces the near-certainty horizon gradient by roughly
    48--88\%, indicating that much of the raw maturity pattern reflects priced
    settlement frictions rather than forecast error alone. Market architecture
    changes the wedge: negRisk conversion compresses discounts by recycling part
    of the position into synthetic collateral, while yield-bearing collateral
    flattens the term structure by reducing the opportunity cost of lock-up. The
    results show that pricing quality in prediction markets is endogenous to
    settlement mechanics, collateral productivity, and capital-recycling design.
    Prediction-market prices therefore aggregate information through a financial
    infrastructure whose funding conditions are measurable and economically
    important.
  \end{abstract}

  \section{Introduction}

  Prediction-market prices are increasingly used as public probability signals.
  Odds from platforms such as Polymarket and Kalshi now appear in political
  commentary, media dashboards, and creator content, often as complements or
  substitutes for polls~\cite{HallKoh2026}. The standard logic is simple: a
  winner-take-all contract pays \$1 if an event occurs and \$0 otherwise, so its
  price can be read as the market-implied probability of the event~\cite{Wolfers2004}.
  Much of the empirical literature adopts this mapping and interprets systematic
  deviations between prices and realized frequencies as informational or
  behavioral error, including miscalibration and favorite--longshot
  bias~\cite{Manski2006,Page2013}.

  This paper argues that the mapping is incomplete for collateralized,
  oracle-settled prediction markets. The standard price-as-probability
  benchmark remains useful, but these venues are information-aggregation
  mechanisms embedded in capital markets rather than frictionless probability
  oracles. Their prices are discounted probabilities: prices of delayed
  contingent claims that combine beliefs with settlement delay, locked-capital
  opportunity cost, liquidity demand, and residual platform or oracle risk.
  Traders allocate collateral to state-contingent claims, and that collateral
  can remain locked even after the real-world uncertainty has effectively
  disappeared. Systematic price--frequency gaps may therefore reflect financial
  architecture as well as forecasting error.

  The intuition is that a nearly certain dollar paid after delayed settlement is
  not the same object as an immediately redeemable dollar. Polymarket's ``Will
  Jesus Christ return in 2025?'' market\footnote{Polymarket market page:
  \url{https://polymarket.com/event/will-jesus-christ-return-in-2025}} is
  illustrative~\cite{Polymarket2026}: for months, the complementary
  near-certain NO side traded around \$0.96. A trader buying \$1{,}000 of NO at
  that price would earn roughly 4.2\% if the position paid \$1 at settlement,
  but only after locking capital for most of the year. A below-par price can
  therefore be consistent with near certainty once outside returns, liquidity
  needs, and residual platform risk are priced in.

  We formalize this friction as settlement-induced discounting. Let
  $X_i\in\{0,1\}$ denote the settlement payoff of contract $i$ and let
  $\tau_{i,t}$ denote the remaining time until formal resolution. Instead of the
  frictionless benchmark $P_{i,t}=\mathbb{E}_t[X_i]$, we study the reduced-form
  pricing equation
  \[
    P_{i,t}=\mathbb{E}_t[X_i]D(\tau_{i,t}),
  \]
  where $D(\tau)\leq 1$ summarizes the market value of delayed settlement,
  capital lock-up, outside opportunities, liquidity, and residual platform risk.
  Observed prices are therefore discounted probabilities rather than
  frictionless probabilities; interpreting them as probabilities requires
  accounting for settlement discounting. The wedge should be largest at longer
  settlement horizons, should generate apparent
  underconfidence in near-certain contracts, and should shrink when market design
  allows capital to earn yield or be recycled earlier.

  The empirical challenge is that high prices need not represent certainty: they
  may still contain residual belief uncertainty. We therefore use persistent
  upper-tail contracts as candidate support for recovering the discount
  component: markets in which one side remains highly priced for multiple
  consecutive days and does not subsequently reverse. We estimate a high-quantile near-certainty frontier by realized
  time to settlement and convert that frontier into an annualized settlement
  wedge (ASW), a reduced-form required-return measure for locked
  prediction-market capital.

  We find a persistently positive and maturity-dependent ASW term structure.
  Discount adjustment reduces the near-certainty horizon gradient by roughly
  48--88\%, and event-level cross-fitting yields similar attenuation. Under the
  highest-wedge robust frontier considered, the residual gradient becomes
  statistically indistinguishable from zero. These results indicate that a
  substantial share of apparent long-horizon miscalibration reflects mechanical
  settlement discounting rather than purely informational error.

  Relative to prior conceptual work on cost-of-capital channels~\cite{Grant2019}
  and earlier evidence on horizon effects~\cite{Page2013}, our contribution is
  threefold:
  \begin{enumerate}
    \item We model collateralized prediction-market prices as discounted
          expected payoffs under delayed settlement and locked collateral.
    \item We recover a frontier-implied ASW term structure using near-certain
          contracts and realized settlement times, and show that discount
          adjustment attenuates horizon-conditioned calibration distortion.
    \item We show that market design changes the discount wedge: negRisk
          conversion mechanically compresses near-certainty discounts, and
          cross-platform evidence is consistent with yield-bearing collateral
          flattening the wedge.
  \end{enumerate}
  Thus, forecast evaluation, cross-platform comparison, and market-design
  assessment must account for the financing environment in which
  prediction-market prices are formed.

  \section{Background}

  This section isolates the primitives needed for the pricing
  mechanism. The causal chain is simple: economically resolved claims can
  remain formally unsettled; unsettled claims keep capital from being
  redeployed; and market architecture determines whether that lock-up is
  mitigated through conversion or collateral yield.

  \subsection{Settlement Timing and Locked Collateral}

  The wedge starts with a distinction between knowing an outcome and redeeming a
  claim. For each market $i$, let $T_i^{\mathrm{event}}$ denote the time at
  which the outcome is effectively determined from publicly available
  information; $T_i^{\mathrm{end}}$ the platform's scheduled market end or
  trading-close metadata;
  $T_i^{\mathrm{oracle}}$ the time at which the oracle outcome becomes finalized;
  and $T_i^{\mathrm{settle}}$ the first time at which winning claims
  can be redeemed. These timestamps need not coincide: $T_i^{\mathrm{event}}$
  can precede $T_i^{\mathrm{end}}$, and both can precede oracle finalization
  and redemption.

  On Polymarket, resolution is handled through the UMA Optimistic Oracle, under
  which anyone can propose an outcome, undisputed proposals finalize after a
  two-hour challenge window, and disputed cases can escalate to UMA's voting
  mechanism~\cite{Polymarket2026a,UMA2026}. This architecture creates an oracle
  stage between economic resolution and redeemability: even after the outcome is
  clear, claims become cash only when the oracle path finalizes. Empirically, we
  proxy settlement completion using realized oracle-finalization timestamps,
  since redemption becomes possible only after oracle resolution.

  In the framework below, $T_i^{\mathrm{resolution}}$ refers to
  $T_i^{\mathrm{settle}}$, not the scheduled market end. The economic
  implication is direct: capital tied to unresolved claims cannot be redeployed
  until redemption and therefore forgoes outside returns. Claimholders may also
  bear liquidity costs and residual oracle or platform risk during the
  settlement interval. A near-certain \$1 payoff is therefore a delayed claim
  rather than an immediately spendable dollar.

  \subsection{Contract Architecture}

  A binary prediction contract is a state-contingent claim paying \$1 if an
  event occurs and \$0 otherwise. On decentralized venues, such claims are often
  implemented through the Gnosis Conditional Token Framework~\cite{Gnosis2020}
  (CTF): collateral is locked, a complete set of outcome-contingent claims is
  minted, and that set can be split into tradable outcome tokens. A complete set
  sums to par at resolution, while losing outcome tokens expire worthless.

  Only one feature of this architecture is essential for the model: unless the
  position can be merged, converted, or otherwise recycled, the collateral value
  embedded in unresolved outcome tokens remains unavailable until settlement.
  Protocol operations that recombine claims into complete-set or
  collateral-equivalent positions therefore matter because they shorten the
  effective lock-up even when the formal settlement date is unchanged.

  \subsection{Capital-Efficient Market Design}

  Market design changes the wedge by changing either the effective duration of
  lock-up or the return earned while collateral is locked. We use this logic to
  separate ordinary binary markets from designs that mechanically compress the
  settlement discount.

  \paragraph{Conversion.}
  Some Polymarket mutually exclusive multi-outcome events use a negative-risk
  (negRisk) architecture~\cite{Polymarket2023}. In these events, a multi-outcome
  question is represented as a collection of binary outcome markets, one for
  each candidate outcome, and an additional conversion layer links the binary
  markets within the same event. The key mechanism is that a NO share in one
  outcome can be converted into YES shares in the other outcomes. In a
  three-outcome event $(A,B,C)$, for example, converting $N_A+N_B$ is equivalent
  to one residual YES on $C$ plus one cash-like par unit.
  Appendix~\ref{app:neg-risk-identity} formalizes and proves this payoff
  identity. The delayed-settlement
  exposure is therefore concentrated in the smaller residual claim, so we treat
  negRisk as a distinct design feature and analyze it separately from the
  non-negRisk baseline.

  \paragraph{Collateral yield.}
  A second design mechanism is compensation for locked balances. Polymarket
  historically operated with largely non-yield-bearing locked collateral, but it
  later introduced fixed 4\% holding rewards for selected long-dated
  market positions~\cite{Shittu2025,Polymarket2026b}. Kalshi~\cite{Kalshi2026} documents
  APY paid on collateral balances, with policy rates that can change over
  time~\cite{Kalshi2026a}. Yield support does not eliminate delayed settlement,
  but it offsets the opportunity cost of waiting and should flatten the observed
  discount wedge.

  \section{Settlement-Induced Discounting}

  \subsection{Pricing with Delayed Settlement}

  \[
    P_{i,t}=\mathbb{E}_{t}[X_{i}]\cdot D(\tau^{e}_{i,t}), \qquad D(\tau)\in(0,1].
  \]
  Here $X_i\in\{0,1\}$ is the payoff at settlement, $P_{i,t}$ is the time-$t$
  price, and $\tau^{e}_{i,t}$ is the expected remaining time until the winning
  claim becomes redeemable. The standard frictionless benchmark corresponds to
  $D(\tau)=1$. We use ``settlement discounting'' as shorthand for the
  reduced-form pricing wedge induced by delayed redeemability, incorporating
  opportunity cost, residual settlement risk, liquidity frictions, and capital
  lock-up.

  A convenient parameterization is
  \[
    D(\tau)=\exp\!\bigl(-r_{\mathrm{PM}}(\tau)\,\tau\bigr),
  \]
  where $r_{\mathrm{PM}}(\tau)$ is the reduced-form required return on capital
  locked in unresolved prediction-market positions, measured in the same time
  units as $\tau$. It is not a risk-free rate or a pure oracle-delay premium;
  it is the revealed compensation for holding a delayed, platform-specific
  settlement claim. The annualized settlement wedge (ASW) recovered below
  reports this required-return object in annualized terms. Unlike classical
  prediction-market horizon effects, the discount factor here is directly shaped
  by settlement architecture, collateral mobility, and convertibility.

  The priced horizon is ex ante expected remaining settlement time,
  \[
    \tau^{e}_{i,t}:=\mathbb{E}_{t}\!\left[T_{i}^{\mathrm{resolution}}-t\right],
  \]
  where $T_i^{\mathrm{resolution}}$ denotes the priced redeemability timestamp,
  i.e., $T_i^{\mathrm{settle}}$. In the main structural panel, realized
  \texttt{closeDate} proxies this settlement timestamp. Empirically,
  \[
    \tau^{\mathrm{obs}}_{i,t}
    =T_{i}^{\mathrm{resolution}}-t
    =\tau^{e}_{i,t}+u_{i,t},
  \]
  where $u_{i,t}$ is forecast error in settlement timing. Under standard
  non-differential proxy-error conditions, using realized settlement time
  attenuates estimated horizon effects toward zero rather than mechanically
  generating a positive horizon gradient. For notational simplicity, subsequent
  equations write $\tau_{i,t}$.

  \paragraph{Settlement--Discount Continuum.}
  The same discount factor links long-horizon below-par pricing and
  short-horizon post-event carry. At long horizons, the settlement wedge is
  visible in near-certain prices; near settlement, the level wedge is smaller,
  but a few basis points can annualize into high carry because the holding
  period is short. Tick size and discrete order books can prevent the wedge from vanishing completely even after economic
  uncertainty is gone. These are therefore not separate anomalies, but different
  points on one maturity-dependent settlement-discount continuum; Appendix~\ref{app:strategy-evidence}
  documents the corresponding short-horizon carry strategies.

  \subsection{Testable Implications}

  The framework yields three direct empirical predictions when $D'(\tau)<0$.

  \paragraph{(1) Long-Horizon Near-Certainty Discounts.}
  For near-certain contracts, delayed settlement implies below-par pricing:
  \[
    P_{i,t}=\mathbb{E}_{t}[X_{i}]D(\tau_{i,t})<1
    \quad \text{when }\mathbb{E}_{t}[X_{i}]\approx 1,\; \tau_{i,t}>0.
  \]
  Among contracts with similar expected payoffs, longer remaining settlement
  delay implies a lower price. If
  $\mathbb{E}_{t}[X_i]\approx \mathbb{E}_{t}[X_j]\approx 1$ and
  $\tau_{i,t}>\tau_{j,t}$, then
  \[
    P_{i,t}\approx D(\tau_{i,t})<D(\tau_{j,t})\approx P_{j,t}.
  \]
  Interpreting raw prices as probabilities therefore generates a horizon-linked
  underconfidence pattern, which discount adjustment should attenuate.

  \paragraph{(2) Mechanical Price Drift.}
  Conditional on beliefs,
  \[
    \frac{\partial P_{i,t}}{\partial \tau_{i,t}}
    =\mathbb{E}_{t}[X_i]D'(\tau_{i,t})<0.
  \]
  Equivalently, as calendar time passes and $\tau_{i,t}$ falls, prices drift
  upward mechanically absent belief revisions.

  \paragraph{(3) Platform-Design Sensitivity.}
  Platform features operate through $D(\tau)$. Yield-bearing collateral, faster
  settlement, merging, and conversion paths should raise $D(\tau)$ toward one
  for a given horizon and compress the wedge; longer lock-up or lower collateral
  productivity should lower $D(\tau)$ and amplify it. Settlement discounting is
  therefore endogenous to market architecture.

  These predictions motivate the empirical design in the next section: we
  estimate the settlement-discount frontier, recover ASWs across horizons, test
  horizon-conditioned calibration, and compare environments with different
  capital efficiencies.

  \section{Data and Empirical Design}

  This section maps the settlement-discount model to the empirical design. We
  first describe the analysis samples, then define the identification strategy,
  the near-certainty frontier used to estimate the discount factor, and the
  calibration adjustment used in the main tests.

  \subsection{Data Sources and Sample Construction}

  \paragraph{Data sources.}
  The main empirical setting is Polymarket. We use hourly quote data from the
  Polymarket Data API and on-chain trade and participant data from the
  Polygon blockchain~\cite{Polygon2026}. In the notation above, \texttt{endDate}
  measures $T_i^{\mathrm{end}}$ and \texttt{closeDate} is the realized proxy
  for $T_i^{\mathrm{settle}}$ in the structural panel. For appendix trade
  diagnostics, we link UMA oracle timestamps and dispute metadata, with
  \texttt{umaEndDate} as the realized oracle-resolution field. External
  opportunity-cost benchmarks are AAVE lending rates from Dune~\cite{DuneAAVE2026}
  and the 2-year constant-maturity Treasury yield from FRED~\cite{FREDDGS22026}.

  \paragraph{Sample construction.}
  The raw universe contains all Polymarket markets listed from platform
  inception through December 31, 2025. The structural sample restricts this
  universe to markets where delayed settlement can plausibly matter: event
  duration at least 14 days, cumulative outcome-token volume
  \texttt{volumeNum}$\ge 100$, attached CLOB price histories, and non-stale
  daily quote panels. Within this universe, we focus on markets where it is
  meaningful to study near-certain pricing: at least one YES or NO midpoint must
  remain at or above $0.90$ for seven consecutive daily snapshots, and markets
  with subsequent large reversals are excluded. Baseline frontier and
  calibration exercises exclude negRisk markets to estimate a non-negRisk
  discount benchmark; negRisk markets re-enter only in the design-comparison
  exercise. Detailed stage-level sample construction is reported in
  Appendix~\ref{app:etl-stage-accounting}, Table~\ref{tab:appendix_pipeline_coverage}.

  The structural panel supports frontier estimation, ASW recovery, and
  horizon-conditioned calibration. Calibration uses one non-negRisk observation
  per market-day at 00:00 ET, so raw and discount-adjusted diagnostics share
  the same pricing architecture as the frontier used to estimate
  $\hat D(\tau)$. A separate linked trade dataset contains 19{,}868{,}940
  trades and supports the short-horizon carry and address-level appendix
  analyses.

  Table~\ref{tab:analysis-sample-map} summarizes the analysis samples and
  measurement choices.
  
\begin{table}[t]
  \centering
  \caption{Main analysis samples and measurement choices.}
  \label{tab:analysis-sample-map}
  \resizebox{\linewidth}{!}{%
  \begin{tabular}{llllll}
    \hline
    \textbf{Exercise} & \textbf{Sample} & \textbf{Unit} & \textbf{negRisk} & \textbf{Horizon} & \textbf{Use} \\
    \hline
    Structural panel & Structural panel & Market-day & Mixed & \texttt{closeDate} & Base sample \\
    Baseline frontier & Near-certainty tail & Horizon bin & Excluded & \texttt{closeDate} & Estimate $\hat D(\tau)$ \\
    Calibration & Non-negRisk tail & Market-day & Excluded & \texttt{closeDate} & Calibration tests \\
    negRisk comparison & Named-outcome negRisk & Weekly & Included & Weeks-to-end & Conversion effect \\
    Kalshi comparison & Cross-platform tail & Weekly & PM baseline & Weeks-to-end & Yield comparison \\
    Carry appendix & Linked trades & Trade & Mixed & Trade-to-settle & Post-event carry \\
    \hline
  \end{tabular}%
  }
\end{table}

  \subsection{Identification Strategy}
  \label{sec:discount-identification}

  The central empirical move is to use persistent upper-tail contracts to
  recover the discount component in
  \[
    P_{i,t}=\mathbb{E}_{t}[X_{i}]D(\tau_{i,t}).
  \]
  Define residual failure probability
  \[
    \delta_{i,t}:=1-\mathbb{E}_{t}[X_{i}]\in[0,1],
  \]
  so that
  \[
    P_{i,t}=(1-\delta_{i,t})D(\tau_{i,t}).
  \]
  Rearranging the pricing identity isolates the confound:
  \[
    1 - P_{i,t}
    =
    \underbrace{1 - \mathbb{E}_{t}[X_{i}]}_{\text{residual uncertainty}}
    +
    \underbrace{\mathbb{E}_{t}[X_{i}]\bigl(1-D(\tau_{i,t})\bigr)}_{\text{settlement discount}}.
  \]
  Distance from par therefore combines residual uncertainty and settlement
  discounting. The main identification threat is horizon-correlated residual
  uncertainty: a long-horizon near-certain claim may trade below par because it
  is discounted, because it remains uncertain, or because both forces operate.

  Our main estimates use the \emph{high-price near-certainty sample}: claims
  enter when one side trades persistently in the upper tail. To validate this
  sample, we separately identify a smaller set of economically resolved
  markets, where the outcome is already known but oracle finalization and
  redemption are still pending. In this window,
  $T_i^{\mathrm{event}}\le t<T_i^{\mathrm{settle}}$, residual outcome
  uncertainty is minimal and
  \[
    P_{i,t}\approx D(\tau_{i,t}).
  \]
  Appendix~\ref{app:manual-frontier-validation} reports a compact manual audit
  of this resolved validation sample.

  Identification is exact when $\delta_{i,t}=0$. In the data, the frontier
  interpretation rests on four conditions.

  \begin{enumerate}
    \item \textit{Near-certainty.} In the persistent upper tail, residual
          failure probability is small relative to the price--par wedge.
    \item \textit{Frontier selection.} Conditional on horizon, the upper
          envelope is supported by observations with the least residual
          uncertainty and smallest idiosyncratic downward frictions.
    \item \textit{Proxy horizon.} Realized settlement time is a noisy but
          informative proxy for the ex ante priced settlement horizon.
    \item \textit{No matching residual trend.} After the persistence and
          reversal screens, frontier-support residual uncertainty is not rising
          with horizon strongly enough to explain the observed price gradient.
  \end{enumerate}

  Under these conditions, a high-quantile frontier satisfies
  \[
    P_{q}(\tau)\approx (1-\delta_{q}(\tau))D(\tau),
  \]
  where $\delta_q(\tau)$ is residual uncertainty among frontier-support
  observations. If $\delta_q(\tau)>0$, then
  \[
    \hat{D}_{q}(\tau)=P_{q}(\tau)\le D(\tau),
  \]
  and the implied rate is upward biased:
  \[
    \hat{r}_{q}(\tau)=-\frac{1}{\tau}\log P_{q}(\tau)
    =r_{\mathrm{PM}}(\tau)-\frac{1}{\tau}\log\!\bigl(1-\delta_{q}(\tau)\bigr)
    \ge r_{\mathrm{PM}}(\tau).
  \]
  The recovered object is therefore a reduced-form settlement wedge. Remaining
  residual uncertainty biases ASWs upward and makes discount-adjusted
  calibration an attenuation test rather than exact latent-probability recovery.

  \subsection{Frontier Estimation}
  \label{sec:near-certainty-frontier}

  Define the near-certainty frontier $P_q(\tau)$ as the high-quantile envelope
  of retained upper-tail prices at settlement horizon $\tau$. Using a high quantile rather than the raw maximum limits
  sensitivity to stale quotes and single-market outliers.

  The implied continuously compounded lock-up rate is
  \[
    r_{q}(\tau)=-\frac{1}{\tau}\log P_{q}(\tau),
  \]
  with $\tau$ measured in days. We report the annualized settlement wedge
  (ASW) as
  \[
    \mathrm{ASW}_{q}(\tau)=\exp\!\bigl(365\,r_{q}(\tau)\bigr)-1.
  \]
  The curve $\tau\mapsto\mathrm{ASW}_{q}(\tau)$ is a reduced-form
  required-return term structure for locked prediction-market capital.

  To track calendar-time evolution, we define
  \[
    \mathrm{ASW}_{\min}(t)
    =
    \min_{\tau\in T(t)}\mathrm{ASW}_{q,t}(\tau),
  \]
  where $\mathrm{ASW}_{q,t}(\tau)$ denotes the date-$t$ frontier-implied
  annualized settlement wedge at horizon $\tau$, and $T(t)$ is the set of
  observed horizons with sufficient support on date $t$. Separately, to
  measure how concentrated the frontier-support tail is across markets at a
  given settlement horizon, we report
  \[
    N_{\mathrm{eff}}(\tau)=\left(\sum_{i}w_{i}(\tau)^{2}\right)^{-1},
  \]
  where $w_i(\tau)$ is market $i$'s share of observations in the
  frontier-support tail at horizon $\tau$. Operationally, for each horizon bin
  we compute APYs for all retained observations, select the bottom 5\% APY tail,
  count each market's observations in that tail, and normalize these counts to
  sum to one. Thus $N_{\mathrm{eff}}(\tau)$ measures the effective number of
  distinct markets contributing to the low-APY/high-price frontier-support tail.
  This tail is a nearby support
  diagnostic for the frontier rather than necessarily the exact set of
  observations that determines the maximum-price or 0.1-percentile APY line.

  \subsection{Calibration Design}
  \label{sec:calibration-measurement}

  Calibration compares quoted prices with realized frequencies. Each forecast
  observation consists of price $P_{i,t}$, horizon $\tau_{i,t}$, and outcome
  $X_i$, where $P_{i,t}$ is the midpoint and $\tau_{i,t}$ is time to realized
  \texttt{closeDate}. We use one snapshot per market-day to limit serial dependence and
  cluster inference by market. Calibration weights are
  \[
    w_i = \sqrt{\max(\mathrm{volume}_i,1)},
  \]
  where $\mathrm{volume}_i$ is outcome-token volume. All calibration exercises
  use non-negRisk markets so the pricing architecture matches the frontier used
  to estimate $\hat D(\tau)$.

  The first calibration object is the raw horizon gradient: whether reliability
  deteriorates with $\tau$ when prices are used directly. Pooling across
  maturities is not innocuous because, under settlement discounting, the same
  raw price can correspond to different expected payoffs at different horizons.
  Because midpoint quotes may be stale or wide in sparse long-horizon markets,
  the calibration diagnostics should be read as price-signal diagnostics.

  The second object is the adjusted horizon gradient. Define
  \[
    \tilde{P}_{i,t}=\min\!\left\{1,\frac{P_{i,t}}{\hat{D}(\tau_{i,t})}\right\},
  \]
  with $\hat D(\tau)$ estimated from the non-negRisk near-certainty frontier.
  The test is whether replacing $P_{i,t}$ with $\tilde P_{i,t}$ attenuates the
  raw horizon gradient. This tests whether the estimated settlement
  component explains the maturity pattern in apparent miscalibration. We report
  both in-sample adjustments and event-level cross-fitted specifications, where
  the frontier is estimated on other events and applied to held-out events.

  \section{Results}

  This section establishes four empirical facts. First, near-certain contracts
  trade below par. Second, the wedge is maturity-dependent and summarized by a
  frontier-implied ASW term structure. Economically, the ASW curve is the
  revealed term structure of compensation required to hold capital in
  delayed-settlement contingent claims. Third, frontier-implied discount curves
  explain much of the raw long-horizon price--frequency gradient. Fourth,
  market architecture shifts the wedge as predicted by capital-efficiency
  mechanisms. We present these facts in turn, then point to possible exploitation strategies
   in Appendix~\ref{app:strategy-evidence}.

  \subsection{Near-Certainty Frontier and Annualized Settlement Wedge}

  The near-certainty frontier turns below-par prices into a lock-up compensation
  object. ASW is the reduced-form required return for locked prediction-market
  capital, bundling opportunity cost, settlement latency, liquidity, platform
  risk, and residual uncertainty that survives the frontier filters. We document
  it through the cross-sectional term structure, the lower-tail time series, and
  external-yield comparisons.

\begin{figure}[h]
    \centering
    \includegraphics[width=\textwidth]{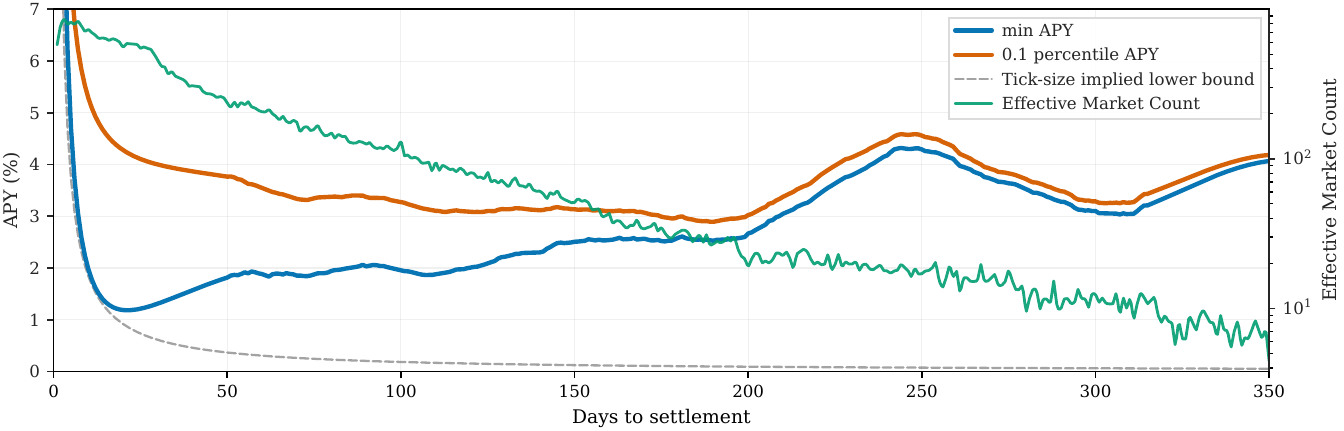}
    \caption{Frontier-implied ASW by time to settlement. The figure plots smoothed minimum and 0.1-percentile ASW across daily contract-level observations. The object is a reduced-form required return on locked prediction-market capital. The dashed line marks the CLOB tick-size lower bound of $0.9995$.}
    \label{fig:apy-yield-curve}
  \end{figure}

  \paragraph{Baseline term structure.}
  As defined in Section~\ref{sec:near-certainty-frontier}, the baseline term
  structure maps the near-certainty frontier into ASW by time-to-settlement.
  Figure~\ref{fig:apy-yield-curve} plots the minimum ASW at each horizon and the
  0.1-percentile ASW as a lower-tail benchmark. Three facts stand out. First,
  both series are positive throughout. Second, ASW falls sharply from the short end, stabilizes by roughly
  20 days, then rises again, with a long-end hump around 230--260 days. Third,
  the short end is partly constrained by CLOB tick size: a $1.0$ ask and
  $0.999$ bid imply a $0.9995$ midpoint. The empirical series remain above that
  bound, so the pattern is not merely mechanical. Because hours or days remain
  at the short end, tiny near-par discounts annualize into high ASWs.
  Appendix~\ref{app:strategy-evidence} links this wedge to post-event carry
  strategies with $0.999$-style bids held through formal settlement. The curve
  therefore implies positive, maturity-varying compensation, with elevated
  annualized rates at the short end and a renewed increase at longer lock-up
  horizons.

\begin{figure}[h]
    \centering
    \includegraphics[width=\textwidth]{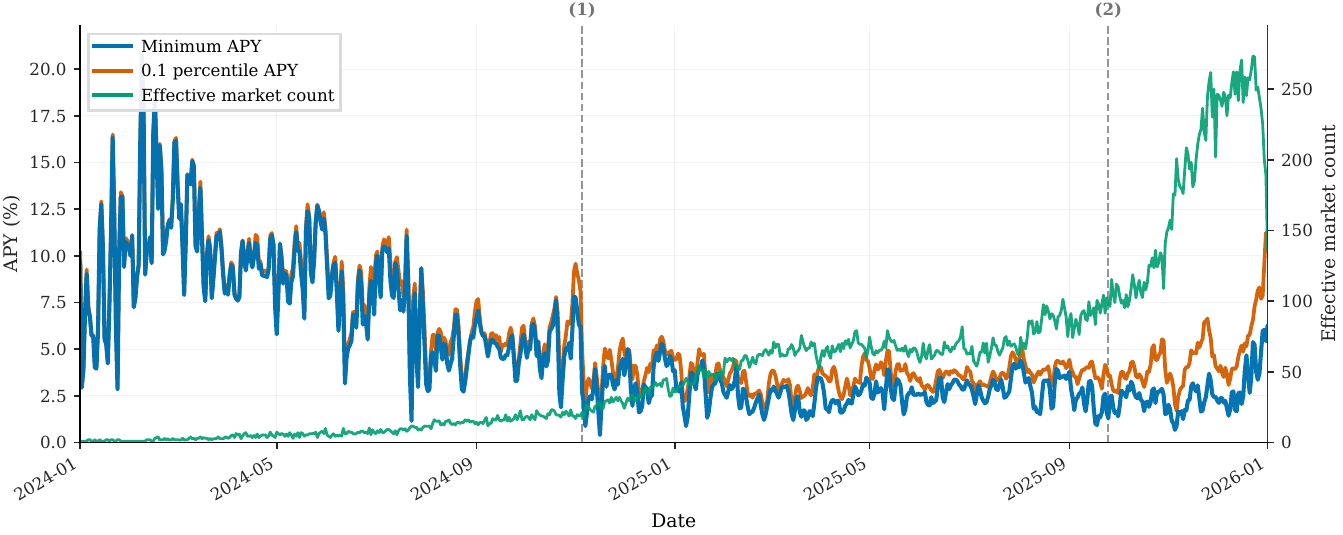}
    \caption{Time-series evolution of lower-tail frontier-implied ASWs with horizon-binned effective support. The figure shows smoothed minimum ASW, 0.1-percentile ASW, and the observation-weighted effective market count $N_{\mathrm{eff}}(\tau)$ for the bottom 5\% APY frontier-support tail. Because $N_{\mathrm{eff}}(\tau)$ is computed by time-to-settlement bin rather than calendar date, it should be read as a concentration diagnostic for the displayed frontier-support tail, not as a pure time-series count of unique markets. Solid vertical lines mark (1) the November 5, 2024 U.S. election and (2) the introduction of Polymarket's 4\% yield program.}
    \label{fig:apy-over-time}
  \end{figure}

  \paragraph{Time-series evolution of the platform frontier.}
  The platform frontier is summarized by the daily lower-envelope statistic
  $\mathrm{ASW}_{\min}(t)$, the lower-tail 0.1-percentile ASW, and the
  horizon-binned support diagnostic $N_{\mathrm{eff}}(\tau)$.
  Figure~\ref{fig:apy-over-time} shows high lower-tail
  ASWs early in the sample, when the relevant frontier-support tail is
  concentrated, consistent with scarce capital. From late 2024 onward,
  frontier-implied rates compress and the relevant horizon bins show broader
  effective support: the frontier becomes cheaper and less concentrated.
  The joint movement of the minimum and lower-tail percentile rules out a
  single-observation artifact. Polymarket's later 4\% yield support for
  long-running markets does not produce a comparable upward break. The late
  sample instead reflects a shift toward shorter horizons, changing the economic
  content of the daily minimum-rate object.

\begin{figure}[h]
    \centering
    \includegraphics[width=\textwidth]{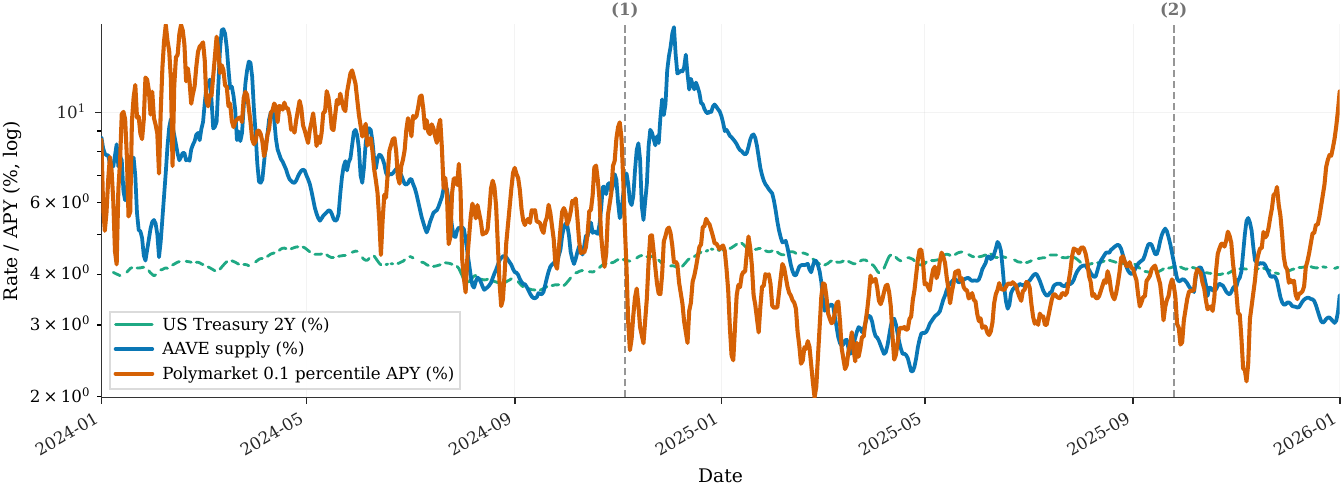}
    \caption{Comparison of external benchmark yields with Polymarket's lower-tail frontier-implied ASW. The figure plots smoothed AAVE supply rates, 2-year U.S. Treasury yields, and Polymarket's 0.1-percentile ASW.}
    \label{fig:apy-yield-comparison}
  \end{figure}

  \paragraph{External benchmark comparison.}
  Figure~\ref{fig:apy-yield-comparison} compares Polymarket's frontier-implied
  ASW with external yields. The relationship is state- and maturity-dependent,
  so the full-period AAVE correlation ($r\approx 0.365$) masks distinct regimes.
  Before the election, when frontier maturities are longer (median about 144
  days), constant-maturity ASWs co-move positively with AAVE (90-day $r=0.302$,
  180-day $r=0.177$). Around the November 5, 2024 U.S. election, the AAVE link
  weakens, consistent with broader participation by less crypto-native traders.
  Between the election and Polymarket's 4\% yield program, ASWs compress and
  benchmark linkage weakens as support thickens. Afterward, frontier maturity
  shifts to the short end (median about 38.5 days), so apparent sign flips
  mainly reflect composition and microstructure, not a reversal in long-horizon
  capital pricing. Treasury correlations stay weak, and lead--lag tests show no
  significant directional relationship with AAVE yields. Appendix~\ref{app:aave-polymarket-scatter}
  shows the corresponding unsmoothed daily scatter.

  These results establish a positive, horizon-varying settlement discount in the
  data. The next subsection asks how much of the raw calibration-horizon pattern
  survives after discount adjustment.

  \subsection{Horizon-Conditioned Price-Signal Distortion}

  Using the settlement discount above and the design in
  Section~\ref{sec:calibration-measurement}, we test whether the raw
  price--frequency gap steepens with settlement horizon and attenuates after
  discount adjustment. All estimates use the non-negRisk structural sample,
  daily market-level observations, and market-clustered inference.

\begin{figure}[h]
    \centering
    \begin{subfigure}[t]{0.49\textwidth}
      \centering
      \includegraphics[width=\linewidth]{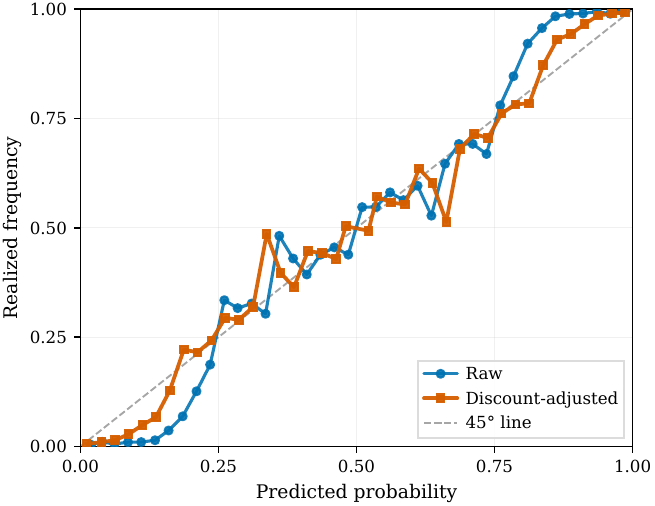}
      \caption{Raw and discount-adjusted reliability for contracts with settlement horizons above 180 days. Observations are volume-weighted; adjusted prices use the date-varying envelope discount $\hat D(\tau)$.}
      \label{fig:discount_adjusted_reliability}
    \end{subfigure}
    \hfill
    \begin{subfigure}[t]{0.49\textwidth}
      \centering
      \includegraphics[width=\linewidth]{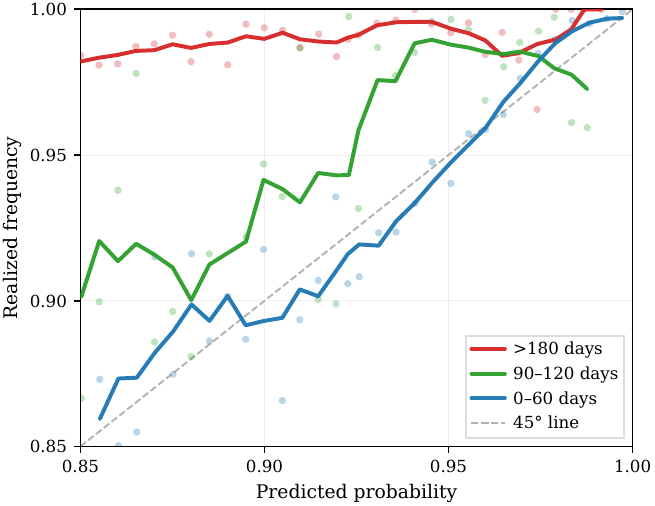}
      \caption{Horizon-conditioned reliability for raw near-certainty prices across settlement-horizon buckets. Observations are volume-weighted; curves are rolling-mean smooths.}
      \label{fig:horizon_conditioned_reliability}
    \end{subfigure}  
    \caption{Calibration diagnostics in the near-certainty region using the same non-negRisk dataset, daily-midpoint sampling, and volume weighting. Panel~(a) shows that discount adjustment moves long-horizon reliability toward the 45-degree line. Panel~(b) shows that raw reliability deteriorates with settlement horizon, matching settlement-induced discounting.}
    \label{fig:reliability_plots}
  \end{figure}

  \paragraph{Reliability Diagnostics.}
  Figure~\ref{fig:reliability_plots} separates raw and adjusted reliability. In
  panel~(b), raw longer-horizon contracts sit progressively above the 45-degree
  line, indicating stronger underprediction as time-to-settlement rises. In
  panel~(a), discount-adjusted long-horizon reliability based on $\hat D(\tau)$
  shifts materially toward the diagonal. The adjustment absorbs much of the
  long-horizon price-signal distortion, especially where the raw settlement
  wedge is largest.

  Table~\ref{tab:calibration_summary} gives the horizon-bucket summary. Raw
  Brier score deteriorates monotonically with horizon (0.0323 in the 0--30 day
  bucket versus 0.1070 beyond 180 days), so raw prices appear less accurate at
  longer maturities. Discount adjustment improves Brier score only modestly,
  because Brier score mixes calibration error with residual outcome uncertainty.
  The weighted calibration-gap MSE more directly isolates deviations from the
  reliability diagonal: improvements rise from 1.15\% at short horizons to
  20.02\% in the longest bucket. Appendix~\ref{app:calibration-summary-plot}
  visualizes this pattern.
  The key diagnostic is the maturity gradient: adjustment matters most where
  capital remains locked longest.

  \begin{table}[t]
    \centering
    \caption{Calibration summary for raw and discount-adjusted prices across
    disjoint settlement-horizon buckets. Brier score measures overall forecast
    error; MSE\textsuperscript{a} measures deviation from the reliability
    diagonal and is the primary calibration diagnostic. Percentage improvements
    are relative to the raw metric.}
    \label{tab:calibration_summary}
    \resizebox{\linewidth}{!}{%
    \begin{tabular}{lrrrrrrr}
      \hline
      \textbf{Horizon} & \textbf{N} & \textbf{Brier (raw)} & \textbf{Brier (adj)} & \textbf{\% imp.} & \textbf{MSE\textsuperscript{a} (raw)} & \textbf{MSE\textsuperscript{a} (adj)} & \textbf{\% imp.} \\
      \hline
      0--30 days       & 194406     & 0.032310             & 0.032259             & 0.16\%           & 0.003102                   & 0.003066                   & 1.15\%           \\
      30--60 days      & 109123     & 0.063199             & 0.063031             & 0.27\%           & 0.004041                   & 0.003831                   & 5.20\%           \\
      60--120 days     & 111181     & 0.084946             & 0.084655             & 0.34\%           & 0.003255                   & 0.002865                   & 11.99\%          \\
      120--180 days    & 50769      & 0.085622             & 0.084938             & 0.80\%           & 0.005903                   & 0.005183                   & 12.18\%          \\
      $>180$ days      & 41933      & 0.106967             & 0.106310             & 0.61\%           & 0.004916                   & 0.003932                   & 20.02\%          \\
      \hline
    \end{tabular}%
    }
    \vspace{0.25em}
    \parbox{\linewidth}{\footnotesize \textsuperscript{a} MSE denotes weighted calibration-gap mean squared error, i.e. squared deviations from the reliability diagonal.}
  \end{table}

  \paragraph{Regression Evidence.}
  As a formal complement, we estimate the horizon gradient in the pricing wedge.
  The goal is to summarize the graphical pattern in one coefficient and test
  whether it survives discount adjustment using frontier-implied curves:
  \[
    \begin{aligned}
      X_i-P_{i,t}&=\alpha+\beta\tau_{i,t}+\varepsilon_{i,t},\\
      X_i-\tilde P^{(q)}_{i,t}&=\alpha_q+\beta_q\tau_{i,t}+\varepsilon_{i,t},
      \qquad
      \tilde P^{(q)}_{i,t}:=\min\!\left\{1,\frac{P_{i,t}}{\hat D_q(\tau_{i,t})}\right\},
    \end{aligned}
  \]
  for frontier choices $q\in\{\mathrm{min},0.1,0.5\}$. All specifications use the
  near-certainty sample ($P\ge 0.9$) and market-clustered standard errors. Here
  $\tau_{i,t}$ proxies for the latent expected horizon priced ex ante, so
  non-differential proxy noise attenuates $\hat\beta$ toward zero.
  Table~\ref{tab:tail_regression_main} shows a positive, highly significant raw
  horizon gradient ($\hat\beta=0.00016780$, $p=3.75\times10^{-7}$). The
  date-varying frontier-implied curves then attenuate it monotonically. The
  minimum-ASW frontier (mean ASW 3.06\%) lowers the slope to 0.00008757 (47.8\%,
  $p=0.0079$); the 0.1-percentile frontier (4.36\%) lowers it to 0.00006686
  (60.2\%, $p=0.0425$); and the conservative 0.5-percentile frontier (6.89\%)
  lowers it to 0.00002076 (87.6\%, $p=0.5285$), statistically indistinguishable
  from zero. Thus discount adjustment reduces the horizon gradient by roughly
  48--88\%, depending on the lower-tail frontier;
  Appendix~\ref{app:maturity-gradient-attenuation} provides the corresponding
  visualization.

  Event-level 5-fold cross-fitting in
  Appendix~\ref{app:crossfit-discount-robustness} tests whether the frontier is
  portable across event sets: each held-out fold is adjusted using frontiers
  estimated from other events. Held-out slope reductions remain large
  (56.4--92.9\%), so attenuation does not weaken when adjusted events are
  excluded from frontier estimation.

  \begin{table}[t]
    \centering
    \caption{Tail regression of the pricing wedge on time-to-settlement in the
    near-certainty sample $P \ge 0.9$. P-values are clustered by market.
    Adjusted specifications use $\tilde P^{(q)}_{i,t}$ constructed from
    date-varying frontier-implied discount curves estimated from the minimum,
    0.1-percentile, and 0.5-percentile near-certainty envelopes. Mean implied
    ASW reports the average ASW along the corresponding frontier curve. The
    sample contains 154{,}298 observations across 4{,}273 market clusters.}
    \label{tab:tail_regression_main}
    \resizebox{\linewidth}{!}{%
    \begin{tabular}{lrrrr}
      \hline
      \textbf{Discount curve} & \textbf{Mean ASW} & $\hat\beta$ & \textbf{Clustered p-value} & \textbf{Slope reduction} \\
      \hline
      Raw (no adjustment)     & ---    & 0.00016780 & $3.75\times10^{-7}$ & ---    \\
      Minimum ASW frontier    & 3.06\% & 0.00008757 & 0.0079               & 47.8\% \\
      0.1-percentile frontier & 4.36\% & 0.00006686 & 0.0425             & 60.2\% \\
      0.5-percentile frontier & 6.89\% & 0.00002076 & 0.5285             & 87.6\% \\
      \hline
    \end{tabular}
    }
  \end{table}

  Calibration and regression evidence point to the same conclusion: the horizon
  effect is economically consistent with discounting. Frontier-implied curves,
  not chosen to eliminate the slope, reduce the raw maturity gradient by roughly
  48--88\%; event-level cross-fitting preserves the ordering. A substantial
  share of the long-horizon price--frequency gap therefore reflects
  settlement-induced pricing of locked capital, not only belief error.

  \subsection{Design-Driven Comparative Statics}

  Market design shapes the level and slope of settlement discounting.
  Convertibility and collateral yield determine how quickly capital can be
  recycled and how costly waiting is. We therefore treat design differences as
  comparative statics of the pricing equation, not shifts in trader rationality.

  \subsubsection{NegRisk Conversion and Discount Compression}

  NegRisk provides the cleanest design comparative static. The key point is
  that a basket of NO tokens can be converted into a deterministic cash-like
  component plus the complementary YES basket. For any subset $S$ of $m$
  outcomes in an $n$-outcome event, Appendix~\ref{app:neg-risk-identity}
  derives the payoff identity
  \[
    \sum_{k\in S} N_k \equiv (m-1)\cdot \mathbf{1}+
    \sum_{j\notin S}Y_j.
  \]
  This payoff identity underlies Polymarket's NegRiskAdapter. The deployed
  adapter exposes the NO-basket-to-complementary-YES conversion direction; at
  the payoff level, the identity says that the NO basket is economically
  equivalent to $(m-1)$ par units plus only the residual complementary YES
  exposure. Thus conversion concentrates delayed-settlement risk in the
  residual leg rather than in the full NO basket.
  Let $V_S(t)$ denote the time-$t$ value of this NO basket. Since the
  deterministic component is valued at par,
  \[
    V_S(t):=\sum_{k\in S}P(N_k,t)
    =(m-1)+\sum_{j\notin S}P(Y_j,t).
  \]
  Under the reduced-form model $P(Y_j,t)=p_j(t)D(\tau)$, with
  $\sum_{j=1}^{n}p_j(t)=1$, this becomes
  \[
    V_S(t)=(m-1)+\bigl(1-p_S(t)\bigr)D(\tau),
    \qquad p_S(t):=\sum_{k\in S}p_k(t).
  \]
  The empirically relevant case is a tail NO basket: the outcomes in $S$ have
  been pushed close to elimination, so their combined YES probability satisfies
  $p_S(t)\approx0$. Then $V_S(t)\approx(m-1)+D(\tau)$. If the $m$ NO tokens are
  similarly priced, their average price
  $\bar P_N(\tau;m):=m^{-1}\sum_{k\in S}P(N_k,t)$ is
  \[
    \bar P_N(\tau;m)
    \approx \frac{(m-1)+D(\tau)}{m}
    =1-\frac{1-D(\tau)}{m}.
  \]
  Thus conversion divides the standard binary settlement wedge $1-D(\tau)$
  across the $m$ nettable NO legs. In the largest nettable basket of an
  $n$-outcome event, $m=n-1$, leaving only one residual YES claim and giving the
  near-certain negRisk benchmark
  \[
    \bar P_N(\tau;n-1)\approx 1-\frac{1-D(\tau)}{n-1}.
  \]
  Appendix~\ref{app:neg-risk-identity} gives the full payoff derivation and
  Appendix~\ref{app:neg-risk-fees} derives fee-adjusted bounds.

\begin{figure}[H]
    \centering

    \begin{minipage}[t]{0.49\textwidth}
      \centering
      \includegraphics[width=\textwidth]{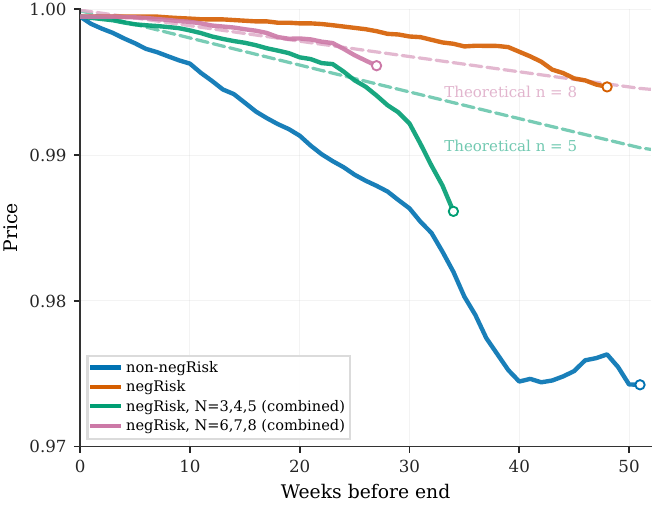}
      \caption{Empirical 99.9th-percentile price frontier by weeks before end.
      Solid lines show trailing moving averages of weekly horizon aggregates
      and are truncated where support falls below 50 markets. Dashed lines
      show conversion-implied benchmarks $\bar P_N(\tau;n)$ for $n=5$ and
      $n=8$ under a constant 4\% discount factor.}
      \label{fig:neg-risk-empirics}
    \end{minipage}
    \hfill
      \begin{minipage}[t]{0.49\textwidth}
      \centering
      \includegraphics[width=\textwidth]{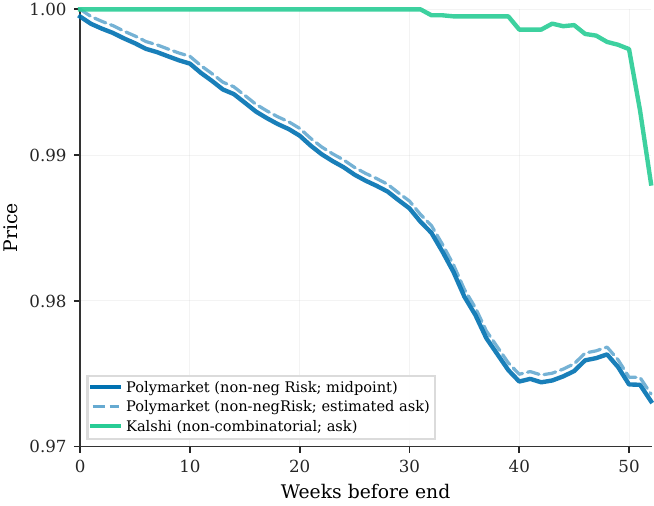}
      \caption{Empirical 99.9th-percentile price frontier by weeks before end. The zoomed near-par scale highlights a cross-platform comparative static: Kalshi remains closer to 1 over most horizons, consistent with flatter effective lock-up premia.}
      \label{fig:kalshi-comparison}
    \end{minipage}
  \end{figure}

  \paragraph{Empirical results.}
  Figure~\ref{fig:neg-risk-empirics} compares the empirical upper envelope of
  near-certain prices with conversion-implied negRisk support lines. The
  envelope uses weekly 99.9th-percentile raw-price frontiers by weeks before
  end. Three patterns match the theory. First, negRisk markets stay closer to
  par than non-negRisk markets across most horizons. Second, within negRisk,
  events with more linked outcomes (e.g., $n=6,7,8$) lie above smaller events
  (e.g., $n=3,4,5$), consistent with stronger netting. Third, the ordering is
  clearest at medium and long horizons, where the wedge is largest; near
  maturity, all series converge toward par. Far-tail subgroup lines are noisier
  as support thins, but the central-horizon ordering remains stable.

  The empirical envelope therefore reproduces the conversion-implied ordering in
  Figure~\ref{fig:neg-risk-empirics}. Economically, negRisk turns delayed
  redemption into a mostly cash-like claim, compressing the settlement wedge by
  construction; compression strengthens with event size. The benchmark treats
  outcome sets as static and tail probabilities as very small; in the appendix we consider
  augmented negRisk-events, non-negligible residual probabilities, conversion fees, and
  taker fees.

  \subsubsection{Yield-Bearing Collateral and Flatter Settlement Premia}

  Yield-bearing collateral provides a second design comparative static. Rather
  than shortening lock-up through conversion, it flattens $\hat D(\tau)$ itself:
  if collateral remains productive while awaiting settlement, delay is less
  costly. Near-certain contracts should trade closer to par and imply a flatter
  ASW term structure.

  Figure~\ref{fig:kalshi-comparison} compares non-negRisk Polymarket contracts
  with Kalshi. Visually, Kalshi's price frontier stays closer to par and is
  flatter across maturities than Polymarket's. Through
  $P_{i,t}=\mathbb{E}_t[X_i]D(\tau_{i,t})$, that flatter near-par price profile
  implies a flatter ASW term structure, which is the comparative static
  predicted when collateral earns yield.

  We interpret the figure as a cross-platform comparative static, not an
  isolated treatment effect: Polymarket and Kalshi differ in users, fees,
  composition, liquidity, collateral rules, and resolution. Read with the
  cleaner negRisk result, however, the comparison reinforces the design
  implication: convertibility and collateral yield compress the wedge through
  distinct capital-efficiency channels.

  \subsection{Trading Exploitation Implications}

  The same term-structure logic has trading strategy implications, documented
  descriptively in Appendix~\ref{app:strategy-evidence}. At the short end, tiny
  pre-settlement discounts can annualize into large \emph{ex post} carry because
  only hours or days remain; at the long end, near-certain contracts can
  temporarily deviate from the platform ASW frontier. Both are retrospective
  exploitations of the same settlement-discount.

  \section{Discussion}

  \paragraph{Core interpretation.}
  Our results suggest that prediction markets should be analyzed as joint
  information-aggregation and capital-allocation systems. A collateralized
  claim is not only a forecast about an event; it is also a claim on a delayed,
  platform-specific dollar. When settlement is delayed and collateral is
  locked, prices embed a funding component tied to time, outside opportunities,
  liquidity, oracle risk, and market design. The standard price-as-probability
  benchmark remains useful, but it is incomplete whenever the discount factor
  varies across maturities or architectures. The central implication is that
  price quality is endogenous to capital efficiency.

  \paragraph{Implications for calibration.}
  Calibration exercises confound belief quality and funding frictions unless
  settlement horizon is controlled for. A long-horizon claim can trade below
  its eventual payoff not because traders doubt the outcome, but because the
  payoff is costly to hold until formal redemption. This mechanism gives a
  direct interpretation to the horizon gradients documented above: discount
  adjustment moves long-maturity near-certainty prices closer to realized
  outcomes because part of the raw price--frequency gap is a settlement wedge.

  Settlement discounting complements rather than replaces behavioral accounts.
  Favorite--longshot bias and other demand-side forces may still affect prices,
  but they operate in an environment where capital is costly to immobilize. The
  mechanism also gives a concise way to interpret some distributional patterns.
  If near-certain claims are discounted, the
  corresponding long-shot side can be mechanically expensive relative to a
  discount-adjusted benchmark, and less sophisticated traders may still buy
  those mechanically overvalued long-shot positions. This is consistent with,
  but does not by itself identify, the profit concentration and long-shot losses
  documented by Akey et al.~\cite{Akey2026}. Conversely, sophisticated participants may earn
  carry-like returns when they can identify claims that are economically safe
  but formally unsettled, subject to liquidity, fees, and resolution-rule risk.

  \paragraph{Capital efficiency as a market-design variable.}
  The most direct design implication is that platforms can improve pricing by
  lowering locked collateral amounts. Conversion is one lever: by shortening
  effective lock-up, negRisk compresses the settlement wedge, especially in
  large outcome sets and when conversion frictions are low~\cite{Polymarket2023}.

  Yield-bearing collateral attacks the same wedge from the other side by
  reducing the opportunity cost of lock-up. Kalshi's APY policy and
  Polymarket's selected holding-rewards program are therefore naturally read as
  market-design responses to long-horizon pricing deterioration
  \cite{Kalshi2026a,Polymarket2026b}. The cross-platform evidence is only
  suggestive, because platforms differ along many dimensions, but it is
  directionally consistent with the model: when collateral or open positions
  earn yield, the implied ASW curve should flatten. For decentralized CTF-style
  markets, non-rebasing yield-bearing collateral provides a clean
  implementation route because it preserves a stable accounting unit while
  allowing yield to accrue in the background; TrueO's TYD is one example of
  this approach~\cite{Trueo2026,Trueo2026a}.

  Polymarket's targeted 4\% rewards also have a possible unintended effect. If
  traders can earn the reward by holding both YES and NO in eligible
  long-running markets, the resulting complete-set position becomes a lower-risk
  platform outside option than marginal exposure to a near-certain frontier
  claim. Rational traders may then require non-rewarded settlement claims to
  clear a higher implied ASW, widening rather than compressing the platform-wide
  wedge. We do not yet see this clearly in recent data, but the mechanism makes
  targeted yield support potentially counterproductive.

  The broader tradeoff is that yield support is visible and easy to understand,
  but can subsidize positions whose settlement wedge is already small. This is
  especially relevant for Polymarket negRisk markets: because conversion already
  recycles capital, subsidizing them is not necessary if the goal is simply to
  offset the cost-of-capital effect. Support is more valuable where collateral
  remains fully locked. Conversion is more targeted because it compresses the
  wedge by construction, but its effectiveness depends on fees. Taker fees and
  conversion fees widen convergence bands, while low-friction conversion makes
  the synthetic collateral channel stronger. From a protocol-design perspective,
  the relevant question is how cheaply the market can recycle capital while
  preserving credible settlement.

  \paragraph{Limits of interpretation.}
  The empirical objects in this paper are reduced-form wedges, not a full
  structural decomposition of required returns. Persistent upper-tail
  prices minimize residual uncertainty, but do not eliminate it; some
  near-certain claims may still contain event risk. Midpoint prices may be
  stale or non-executable in thin markets, especially at longer horizons.
  Realized settlement time is an observable proxy for the expected settlement
  horizon priced by traders at the time of trade. Finally, cross-platform
  comparisons should be read as comparative statics rather than causal
  treatment effects, since platforms differ in users, fees, regulation,
  collateral rules, liquidity, and resolution procedures.

  The broader conclusion is simple: prediction markets are pricing systems
  under collateral constraints. Their prices aggregate information, but they do
  so through contracts that must be funded, held, and eventually redeemed.
  Accounting for that financing environment is therefore necessary for
  calibration, cross-platform comparison, and market-design evaluation.

  \section{Related Work}

  Our paper sits at the intersection of prediction-market interpretation,
  cost-of-capital pricing, and collateral-based valuation. The standard
  price-as-probability mapping remains our benchmark, but delayed settlement and
  locked collateral make the observed price a discounted expected payoff rather
  than a pure probability signal.

  Wolfers and Zitzewitz~\cite{Wolfers2004} establish the canonical
  frictionless benchmark: under risk-neutral, or approximately risk-neutral,
  conditions, winner-take-all prices can be interpreted as market-implied
  probabilities. Manski~\cite{Manski2006} provides a different critique,
  showing that heterogeneous beliefs and budget constraints can prevent market
  price from identifying mean belief. Both papers clarify the belief-to-price
  mapping.

  The closest direct prediction-market predecessor is Page and
  Clemen~\cite{Page2013}. They document calibration distortions and horizon
  effects with explicit attention to time to expiration, opportunity costs, and
  participation/no-trade regions. Their Intrade
  setting makes it difficult to separate event uncertainty, expiration horizon,
  platform costs, and settlement/redeemability frictions. By contrast, realized
  settlement timestamps and persistent near-certainty frontiers allow us to
  recover a discount curve $\hat D(\tau)$ more directly and interpret the
  remaining horizon pattern as a settlement wedge.

  Grant, Johnstone, and Kwon~\cite{Grant2019} provide the closest
  cost-of-capital lens within prediction-market theory. They model prediction
  contracts as risky assets with required returns, producing a belief- and
  investor-specific discount that can rationalize favorite--longshot
  distortions. Our mechanism is complementary but different: the wedge we study
  is infrastructure- and horizon-specific, generated by delayed redemption and
  collateral lock-up. Relative to their framework, our contribution is to
  identify this settlement wedge in field data, measure its term structure, and
  show how market design compresses or amplifies it.

  Outside prediction markets, collateralized-derivatives research provides a
  useful finance anchor. Johannes and Sundaresan~\cite{Johannes2007} show that
  collateralization and mark-to-market conventions alter discounting and
  generate a time-varying cost-of-collateral term structure, while Fujii,
  Shimada, and Takahashi~\cite{Fujii2010} show that collateral remuneration and
  form are not valuation-neutral. The key analogy is empirical as well as
  conceptual: a latent collateral-cost component can be recovered from market
  prices and is economically meaningful. We apply this financing logic to
  prediction markets by recovering a latent settlement wedge from near-certain
  prices with realized oracle-finalization horizons.

  Finally, recent simulation evidence by Maresca~\cite{Maresca2026} provides
  complementary support for the same mechanism. In a controlled market
  experiment, Maresca studies whether remunerating open positions mitigates
  long-horizon distortions and finds that interest-bearing positions improve
  participation and price accuracy. Our contribution is different in both object
  and setting. Rather than estimating the effect of paying interest in a
  designed environment, we use field data to recover the settlement discount
  embedded in observed prices, trace its horizon profile using realized
  settlement time, and compare market architectures with different
  capital-efficiency properties. Maresca therefore provides intervention
  evidence on the effect of remunerating locked capital, whereas we provide a
  valuation-based interpretation of prices under delayed settlement. The
  similarity in direction between Maresca's roughly 83\% mitigation estimate and
  our finding that a conservative 0.5-percentile frontier adjustment shrinks the
  near-certainty horizon gradient by about 88\% is suggestive, but it should not
  be read as an replication, since the underlying estimands differ.

  \section{Conclusion}

  This paper identifies settlement-induced discounting as a measurable pricing
  component in collateralized prediction markets. The standard
  price-as-probability benchmark remains useful, but in these markets observed
  prices are discounted probabilities rather than frictionless probabilities:
  they are prices of delayed contingent claims. When formal redemption is
  delayed, prices combine beliefs about outcomes with settlement delay,
  locked-capital opportunity cost, liquidity demand, and residual platform or
  oracle risk. This component helps account for long-horizon underpricing,
  post-event carry, and maturity-dependent calibration patterns without
  treating them as purely informational or behavioral errors.

  Empirically, we recover settlement-implied discount curves from persistent
  near-certain contracts and convert them into annualized settlement wedges.
  Discount adjustment reduces the near-certainty horizon gradient by roughly
  48--88\%, showing that a large share of the raw maturity pattern reflects a
  priced settlement wedge rather than forecast error alone. The horizon
  structure also links the long and short ends of the market. Distant
  near-certain claims trade below par; near resolution, small residual
  discounts can annualize into large carry.

  Market design is first-order for this pricing component. NegRisk conversion
  and yield-bearing collateral compress the wedge through distinct
  capital-efficiency channels: conversion recovers part of the position as
  synthetic collateral, while yield reduces the opportunity cost of lock-up.
  The cross-design evidence is consistent with these comparative statics, even
  though it should not be read as a standalone treatment effect.

  For researchers and platform designers, the implication is straightforward:
  evaluation frameworks should separate information aggregation from funding
  frictions. Apparent calibration errors can partly reflect financial
  architecture rather than bad forecasting. Calibration and efficiency claims
  should be maturity-conditioned, and interventions should target capital
  efficiency directly---through collateral yield, conversion-friendly market
  structure, and microstructure choices near par. The broader contribution is
  an empirical framework for measuring how settlement mechanics and protocol
  architecture enter prices in modern prediction-market infrastructure.

  \section*{Acknowledgements}

  The complete dataset, including all transient transformations and intermediate
  processing steps, as well as the source code used for the analysis, will be
  published after publication.

  \section{APPENDIX}

  \subsection{Sample Construction and Coverage}
  \label{app:etl-stage-accounting}

  Table~\ref{tab:appendix_pipeline_coverage} summarizes how the empirical
  sample is constructed from the full Polymarket event universe. The table
  separates the broad eligibility screen from the near-certainty screen used
  for identification.

  \begin{table}[H]
    \centering
    \caption{Stage-level sample construction and coverage.}
    \label{tab:appendix_pipeline_coverage}
    \resizebox{\linewidth}{!}{%
    \begin{tabular}{llrrrrr}
      \hline
      \textbf{Stage} & \textbf{Restriction} & \textbf{Events} & \textbf{Markets} & \textbf{negRisk markets} & \textbf{Price obs.} & \textbf{Market-days} \\
      \hline
      Full universe & Full Polymarket crawl & 141{,}848 (100.0\%) & 323{,}342 (100.0\%) & 98{,}186 (30.4\%) & n/a & n/a \\
      Eligible CLOB sample & Volume $>100$, $>2$ weeks, CLOB history & 10{,}608 (7.5\%) & 62{,}291 (19.3\%) & 41{,}439 (66.5\%) & 4{,}647{,}145 & 1{,}473{,}282 \\
      Near-certainty sample & Seven-day upper-tail midpoint screen & 4{,}483 (3.2\%) & 36{,}349 (11.2\%) & 27{,}601 (75.9\%) & 3{,}712{,}775 & 1{,}181{,}226 \\
      \hline
    \end{tabular}%
    }
  \end{table}

  The full universe is collected from active, closed, and archived Gamma event
  documents. The eligible CLOB sample is a coverage and activity screen: it
  keeps markets with positive volume, open intervals longer than two weeks, and
  usable order-book histories. The near-certainty sample then applies the
  identification screen, requiring a YES or NO midpoint of at least $0.90$ for
  seven consecutive daily snapshots from hourly quote data. Price observations
  are valid hourly YES/NO midpoint observations; market-days are distinct
  $(\text{event},\text{market},\text{UTC date})$ observations with positive
  remaining settlement horizon. Event and market percentages are relative to
  the full universe, and negRisk percentages are row shares; the filters raise
  the negRisk share from 30.4\% to 75.9\%, consistent with convertible
  structures being common among longer-duration markets where capital lock-up
  is salient.

  \subsection{Manual Validation of Frontier-Support Observations}
  \label{app:manual-frontier-validation}

  The key identification concern is horizon-correlated residual uncertainty: a
  long-horizon upper-tail contract may trade below par because the outcome is
  not fully certain, not only because settlement is delayed. The main analysis
  addresses this through persistent frontier selection and economically
  resolved but unsettled windows. Table~\ref{tab:manual-frontier-validation}
  below is therefore a plausibility check: it shows that representative
  frontier-support observations involve externally resolvable states where
  residual event risk appears small relative to the observed discount.

  \begin{table}[H]
    \centering
    \caption{Manual validation examples for near-certainty frontier-support observations.}
    \label{tab:manual-frontier-validation}
    \footnotesize
    \begin{tabular}{@{}p{0.28\linewidth}p{0.16\linewidth}p{0.49\linewidth}@{}}
      \hline
      \textbf{Market} & \textbf{Class} & \textbf{Validation rationale} \\
      \hline
      \href{https://polymarket.com/event/how-high-will-inflation-get-in-2025}{U.S. CPI $\geq 8.1\%$ / $10.1\%$ in 2025} & Objective threshold & External BLS threshold; realized inflation was far below the cutoff, so residual uncertainty is small relative to the observed discount. \\
      \href{https://polymarket.com/event/is-earth-flat-2025/is-earth-flat-2025}{Is Earth flat?} & Low-plausibility & Externally resolvable factual proposition; residual uncertainty is plausibly negligible relative to the observed discount. \\
      \href{https://polymarket.com/event/will-jesus-christ-return-in-2025}{Will Jesus Christ return in 2025?} & Low-plausibility & Date-specific low-plausibility proposition; useful as a stress-case diagnostic for non-informational demand at the upper frontier. \\
      \href{https://polymarket.com/event/us-defaults-on-debt-by-2027?tid=1767612065804}{U.S. defaults on debt by 2027?} & Benchmarkable tail event & Externally assessable sovereign credit event; CDS spreads provide an outside benchmark for residual default risk.\footnotemark \\
      \hline
    \end{tabular}
  \end{table}

  \footnotetext{For example, U.S. sovereign CDS spreads during the sample period were on the order of roughly 15 basis points, providing an external market-based proxy for residual default risk~\cite{WGB2026}.}

  \subsection{Supplementary Strategy Evidence}
  \label{app:strategy-evidence}

  This appendix focuses on short-horizon liquidity-demand carry. Once the
  real-world event is economically resolved, a winning claim is close to a
  delayed \$1 payoff. Holders who want immediate balance-sheet release may sell
  slightly below redemption value, while buyers earn carry for warehousing the
  remaining oracle and settlement latency. The price discount can be only a few
  basis points, but annualization over hours or days can be large; the spread
  need not vanish instantly because settlement is non-instantaneous, inventory
  is balance-sheet intensive, and near-par quoting is discretized by tick size.

  The object below is not buying high-probability contracts quoted at
  $0.999$: it is a settlement-liquidity trade in economically resolved claims.
  The next segment defines that screen.

  \subsubsection{Operational Definition and Screen}

  We define \emph{bonding trades} narrowly as post-event settlement-liquidity
  trades: the economic outcome is no longer in doubt, oracle finalization and
  redemption are still pending, and the orderbook contains bids for the winning claim
  at $0.999$ but no asks below par. This is
  not the common Polymarket use of ``bonding'' for buying any high-probability
  share.

  Operationally, we fetch the latest 1{,}500 trades for each market in the full
  Polymarket universe and search for post-event regions with closed status, no
  active dispute, settlement after event end, and a midpoint near $0.9995$.
  Such a midpoint corresponds to bid-side liquidity around $0.999$ and an ask
  effectively at $1.000$: traders cannot buy the claim cheaply from the book,
  but holders can sell certain shares into available bids for immediate cash.
  We retain fills consistent with this sell-into-bid region and compute carry
  from the discount to redemption value over the observed time to settlement.

  \subsubsection{Market Timeline and Post-Event Carry Window}
  \label{app:timeline-trades}

\begin{figure}[H]
    \centering
    \includegraphics[width=\textwidth]{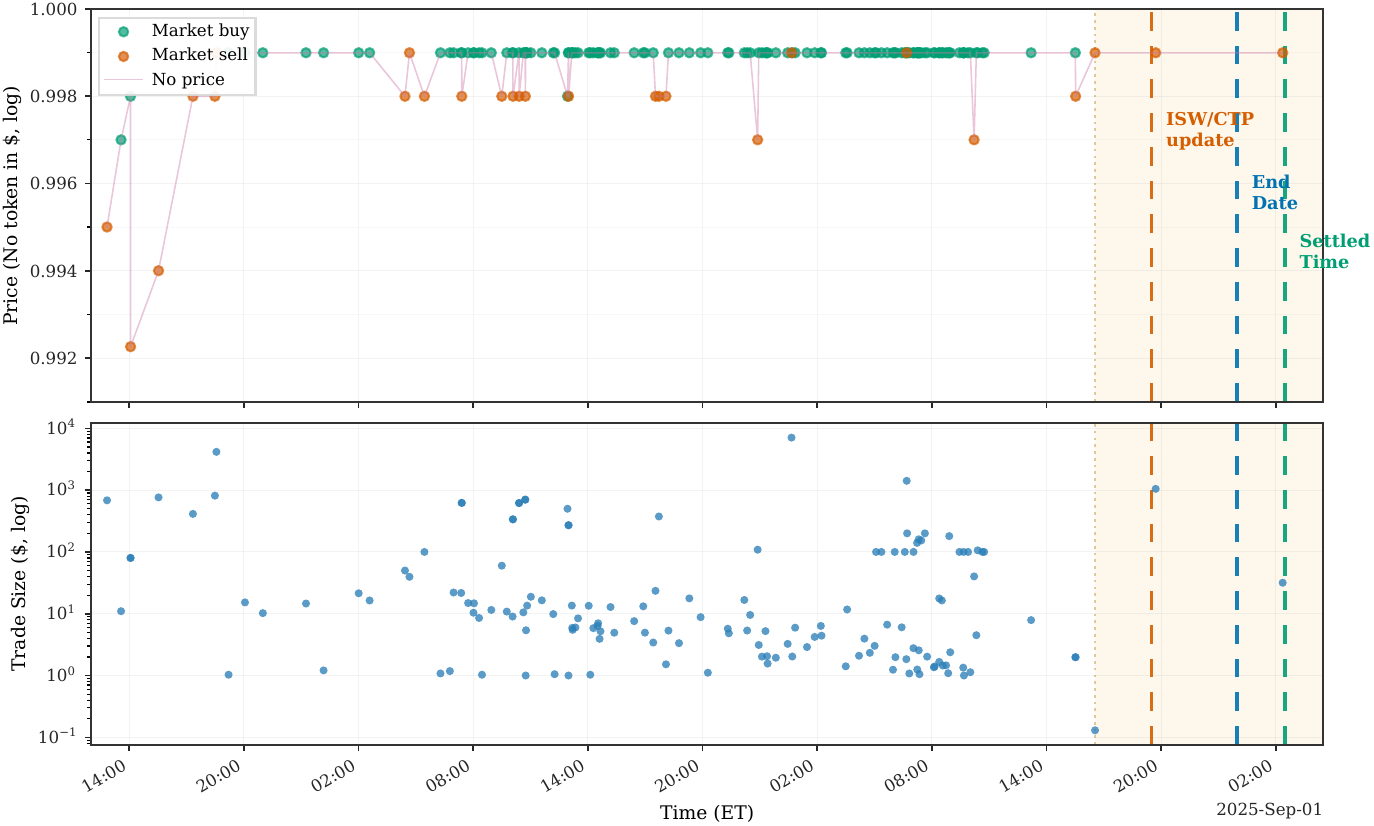}
    \caption{Post-event trading in the Polymarket contract ``Will Russia
    capture Kostyantynivka by August 31?'' The upper panel plots the NO-token
    price around the relevant milestones; the lower panel plots trade size on a
    log scale.}
    \label{fig:market-timeline-trades}
  \end{figure}

  Figure~\ref{fig:market-timeline-trades} presents a case study of
  Polymarket's event ``Will Russia capture
  Kostyantynivka by August 31?''\footnote{Polymarket market page:
  \url{https://polymarket.com/event/will-russia-capture-kostyantynivka-by}.
  The market used the ISW Ukraine map as the primary resolution source, with
  DeepStateMap or credible reporting as fallbacks if needed. A bulletin-board
  clarification required qualifying ISW shading to persist through the next full
  ISW daily update cycle and excluded temporary map glitches and
  infiltration-only shading.} The contract asked whether Russia would capture
  the Kostyantynivka railroad station by August 31, 2025 at 11:59 PM ET. It
  resolved as NO through Polygon UMA Optimistic Oracle V2\footnote{UMA settled oracle record:
  \url{https://oracle.uma.xyz/settled?project=Polymarket&transactionHash=0x3b7eef4440ae5a1c6d7c87a02d32fe460aee93849666b345c33582b97921921e&eventIndex=9}.}
  after a September 1, 2025 proposal and settlement later the same morning.

  The figure aligns trades with the relevant information update, market close,
  and oracle settlement. The shaded region is the case-study carry window: it
  starts at the final transition to a midpoint of $0.9995$ and ends at oracle
  settlement. Before this window, trades still
  reflect ISW battlefield updates and two-sided price discovery. Inside it,
  outcome uncertainty is effectively gone and the book is bid near $0.999$ with
  asks at par: holders can sell certain NO shares into passive bids for
  immediate cash, while liquidity providers warehouse the remaining settlement
  latency.

  \subsubsection{Cross-Sectional Bonding Evidence}
  \label{app:carry-trader-addresses}

\begin{figure}[H]
    \centering
    \includegraphics[width=\textwidth]{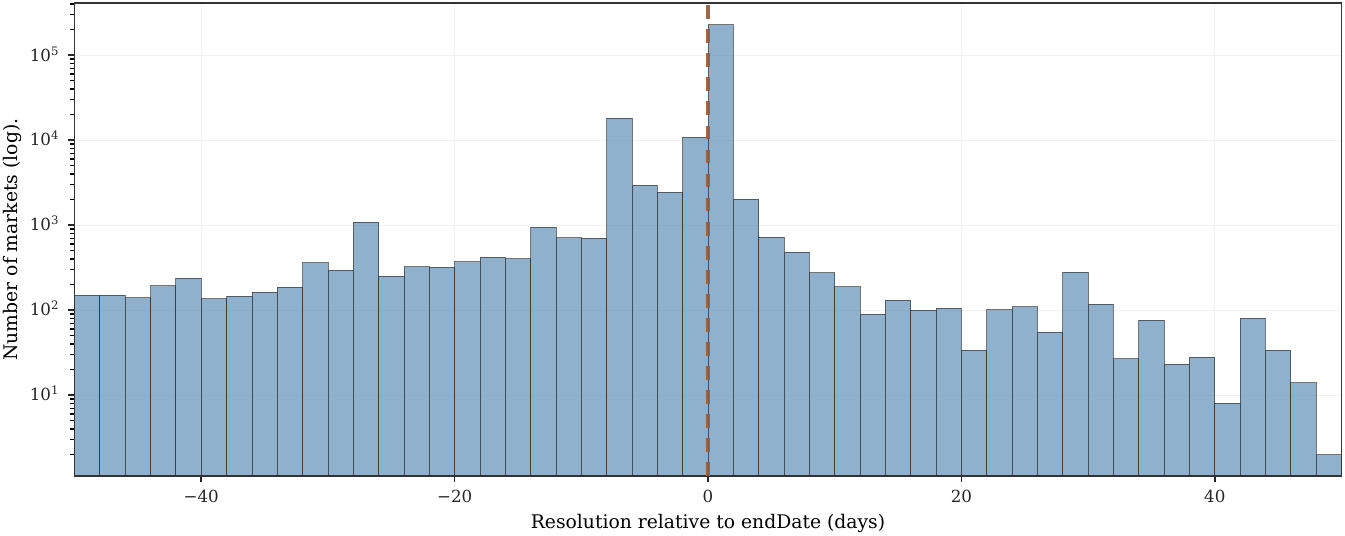}
    \caption{Distribution of remaining settlement horizons for identified
    bonding trades. The figure shows when short-horizon settlement-liquidity
    provision occurs in the screened activity.}
    \label{fig:resolution-time}
  \end{figure}

  The aggregate evidence characterizes the delay between the scheduled EndDate
  and official resolution. This timing is not known ex ante: markets can be
  economically resolved while claims remain non-redeemable, because after
  EndDate they must still wait for oracle finalization. Under the undisputed
  optimistic-oracle path, resolution takes two hours, while disputes can extend
  the interval. Figure~\ref{fig:resolution-time} shows the distribution of
  official resolution times relative to EndDate. Most markets resolve on the
  same day as their EndDate, indicating that the typical post-EndDate wait is
  short even though the exact resolution time is uncertain beforehand.

\begin{figure}[H]
    \centering
    \includegraphics[width=\textwidth]{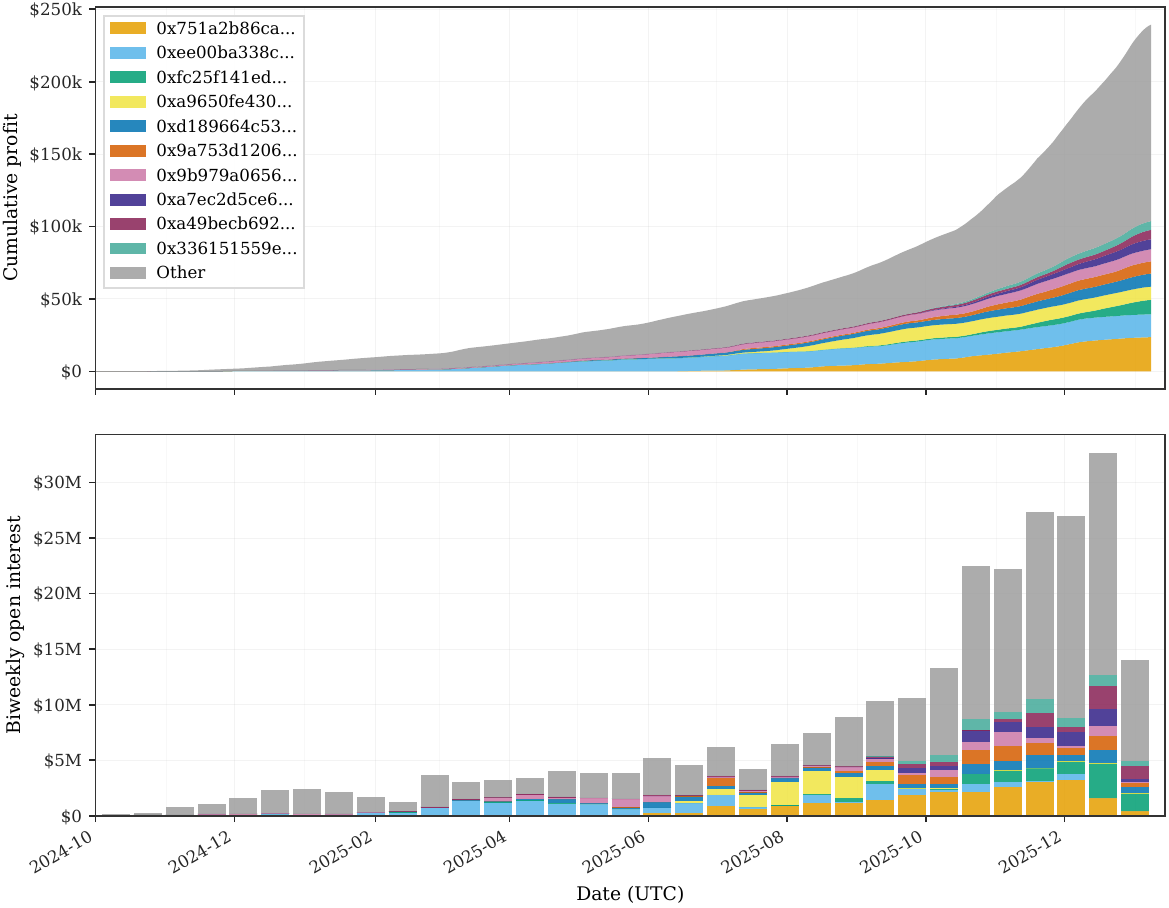}
    \caption{Settlement-liquidity provision in the short-horizon carry screen.
    The top panel shows stacked carry-screen profit; the lower panel shows
    bi-weekly open interest in the screened activity.}
    \label{fig:bonding-statistics}
  \end{figure}

  Figure~\ref{fig:bonding-statistics} shows the aggregate scale of
  settlement-liquidity provision in the short-horizon carry screen. It reports
  both the open interest associated with the screened activity and the profits
  earned by the identified actors over time.

  Table~\ref{tab:carry-trader-addresses} provides the corresponding
  actor-level summary for the largest identified liquidity providers in the
  short-horizon carry screen, sorted by total carry profit. The large
  annualized figures arise mechanically from scaling very small nominal spreads
  over holding periods often measured in hours.

  \begin{table}[H]
    \centering
    \caption{Descriptive, retrospective summary of top short-horizon
    settlement-liquidity providers.}
    \label{tab:carry-trader-addresses}
    \resizebox{\linewidth}{!}{%
    \begin{tabular}{lrrrrr}
      \hline
      \textbf{Address} & \textbf{P\&L (\$)} & \textbf{\# Trades} & \textbf{Median Profit (\%)$^{a}$} & \textbf{Median Profit (\$)} & \textbf{Median hold (hours)} \\
      \hline
      \href{https://polygonscan.com/address/0x751a2b86cab503496efd325c8344e10159349ea1}{\texttt{0x751a}} & 23616.35 & 34994 & 531.67 & 0.040000 & 1.0644 \\
      \href{https://polygonscan.com/address/0xee00ba338c59557141789b127927a55f5cc5cea1}{\texttt{0xee00}} & 15805.16 & 15297 & 428.85 & 0.068490 & 2.0044 \\
      \href{https://polygonscan.com/address/0xfc25f141ed27bb1787338d2c4e7f51e3a15e1f7f}{\texttt{0xfc25}} &  9994.85 &  9649 & 527.45 & 0.039740 & 1.5856 \\
      \href{https://polygonscan.com/address/0xa9650fe4301f45e7f090ada7252f9c1268183565}{\texttt{0xa965}} &  9113.03 & 24440 & 750.11 & 0.047515 & 0.7786 \\
      \href{https://polygonscan.com/address/0xd189664c5308903476f9f079820431e4fd7d06f4}{\texttt{0xd189}} &  9023.79 & 22225 & 565.67 & 0.043040 & 1.0822 \\
      \href{https://polygonscan.com/address/0x9a753d12065a8143ac69ea1732f67daab67b4347}{\texttt{0x9a75}} &  8501.08 &  9613 & 511.37 & 0.042000 & 1.7950 \\
      \href{https://polygonscan.com/address/0x9b979a065641e8cfde3022a30ed2d9415cf55e12}{\texttt{0x9b97}} &  8345.02 & 10340 & 441.72 & 0.058820 & 1.9799 \\
      \href{https://polygonscan.com/address/0xa7ec2d5ce6c38557443a044e627d6abd317279fb}{\texttt{0xa7ec}} &  6924.55 & 11256 & 446.03 & 0.040000 & 1.5369 \\
      \href{https://polygonscan.com/address/0xa49becb692927d455924583b5e3e5788246f4c40}{\texttt{0xa49b}} &  6537.72 &  6089 & 538.01 & 0.041810 & 1.6033 \\
      \href{https://polygonscan.com/address/0x336151559e8c8b048de5231dc8313e196b314363}{\texttt{0x3361}} &  6092.94 & 12988 & 483.59 & 0.040350 & 1.5911 \\
      \hline
    \end{tabular}%
    }
    \parbox{\linewidth}{\footnotesize $^{a}$ Percentage column reports holding-time-adjusted annualized profit; it annualizes small nominal spreads over short holding periods and should not be read as a long-horizon return.}
  \end{table}

\begin{figure}[H]
    \centering
    \begin{subfigure}[t]{0.49\textwidth}
      \centering
      \includegraphics[width=\linewidth]{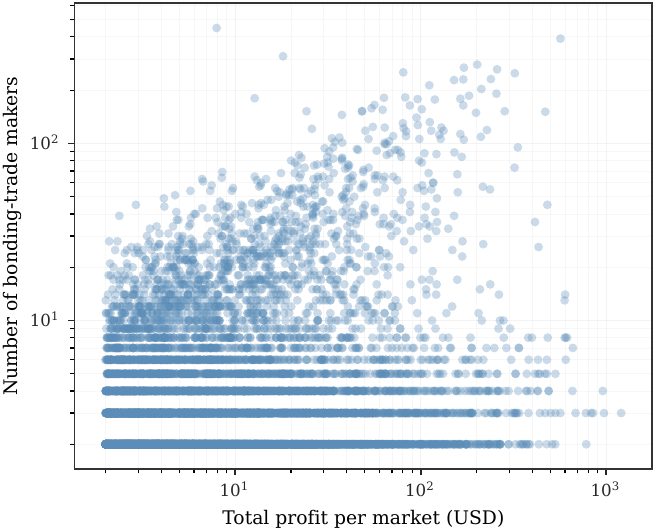}
      \caption{Total profit per market and number of bonding-trade makers for
      the top 100 bonding-trade makers.}
      \label{fig:bonding-scatter}
    \end{subfigure}\hfill
    \begin{subfigure}[t]{0.49\textwidth}
      \centering
      \includegraphics[width=\linewidth]{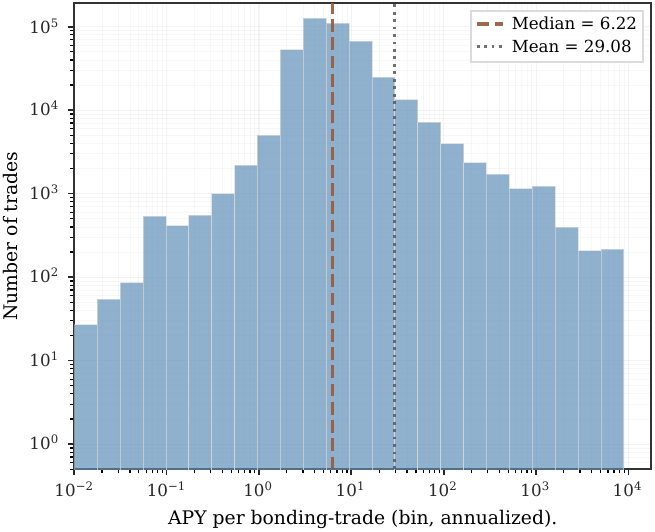}
      \caption{Bonding-trade APY distribution. Histogram bins are APY ranges;
      counts are numbers of trades.}
      \label{fig:bonding-apy-histogram}
    \end{subfigure}
    \caption{Additional descriptive views of settlement-liquidity provision.
    The left panel plots total profit per market against the number of
    bonding-trade makers active in each market. The right panel shows the frequency distribution of bonding-trade
    APYs for the top 100 bonding-traders by transaction-count; the median is 6.22\% and the mean is 29.08\%.}
    \label{fig:bonding-scatter-apy}
  \end{figure}

  Figure~\ref{fig:bonding-scatter-apy} adds two descriptive views of the
  screened bonding trades. The scatter plot relates total profit per market,
  for the top 100 bonding-trade makers, to the number of bonding-trade makers
  active in that market. The histogram reports the distribution of
  bonding-trade APYs: each bin is an APY range and its height counts how many
  trades fall in that range. The median bonding-trade APY is 6.22\%, while the
  mean is 29.08\%.

  \subsection{Supplementary AAVE--ASW Comparison}
  \label{app:aave-polymarket-scatter}

\begin{figure}[H]
    \centering
    \includegraphics[width=0.6\textwidth]{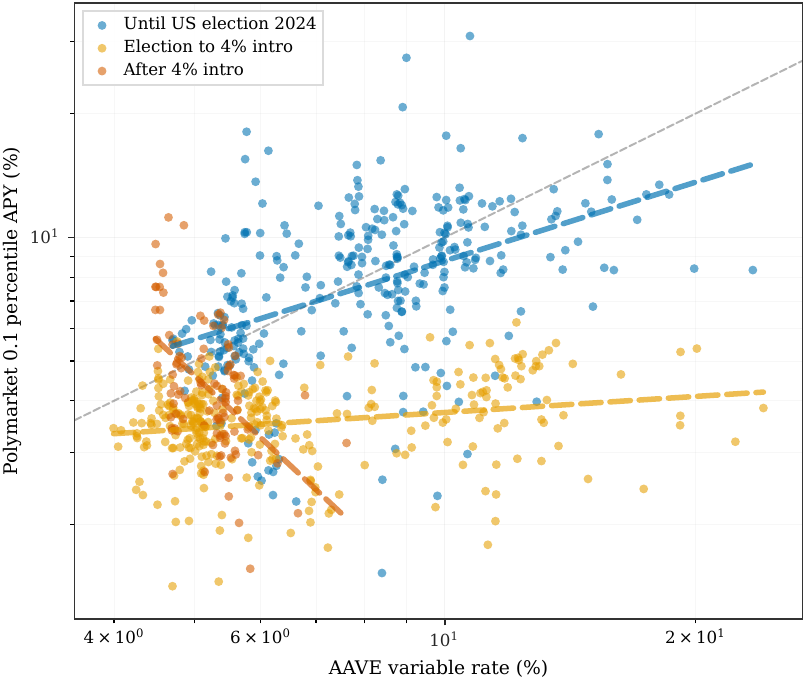}
    \caption{Unsmoothed daily scatter of annualized AAVE variable supply rates and Polymarket daily 0.1-percentile frontier-implied ASWs. Colors distinguish the pre-election period, the post-election/pre-yield-program period, and the period after Polymarket's 4\% yield program. The comparison is descriptive: the two series reflect different liquidity, settlement, and risk structures rather than a direct arbitrage spread.}
    \label{fig:aave-polymarket-scatter}
  \end{figure}

  Figure~\ref{fig:aave-polymarket-scatter} visualizes the unsmoothed daily
  relation between external DeFi yields and Polymarket frontier-implied ASWs
  across market regimes.

  The scatter reinforces the interpretation from the main text: the relation
  between external funding benchmarks and frontier-implied ASWs is noisy,
  regime-dependent, and not reducible to stable arbitrage parity.

  \subsection{Supplementary Calibration Visualization}
  \label{app:calibration-summary-plot}

  \begin{figure}[H]
    \centering
    \includegraphics[width=0.95\textwidth]{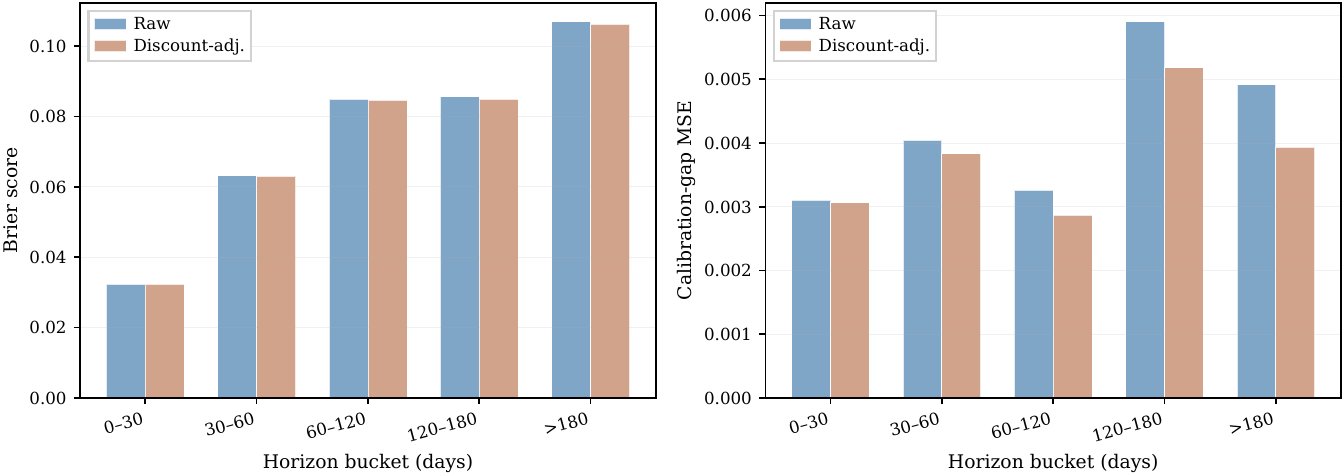}
    \caption{Graphical counterpart to Table~\ref{tab:calibration_summary}. Raw Brier scores worsen with settlement horizon. Discount adjustment changes Brier scores only modestly, but more clearly reduces calibration-gap MSE, especially at longer horizons.}
    \label{fig:appendix-calibration-summary}
  \end{figure}

  Figure~\ref{fig:appendix-calibration-summary} clarifies why calibration-gap
  MSE is the more mechanism-sensitive diagnostic: settlement discounting
  primarily affects reliability, whereas Brier scores also reflect residual
  outcome uncertainty.

  \subsection{Maturity-Gradient Attenuation}
  \label{app:maturity-gradient-attenuation}

  \begin{figure}[H]
    \centering
    \includegraphics[width=0.95\textwidth]{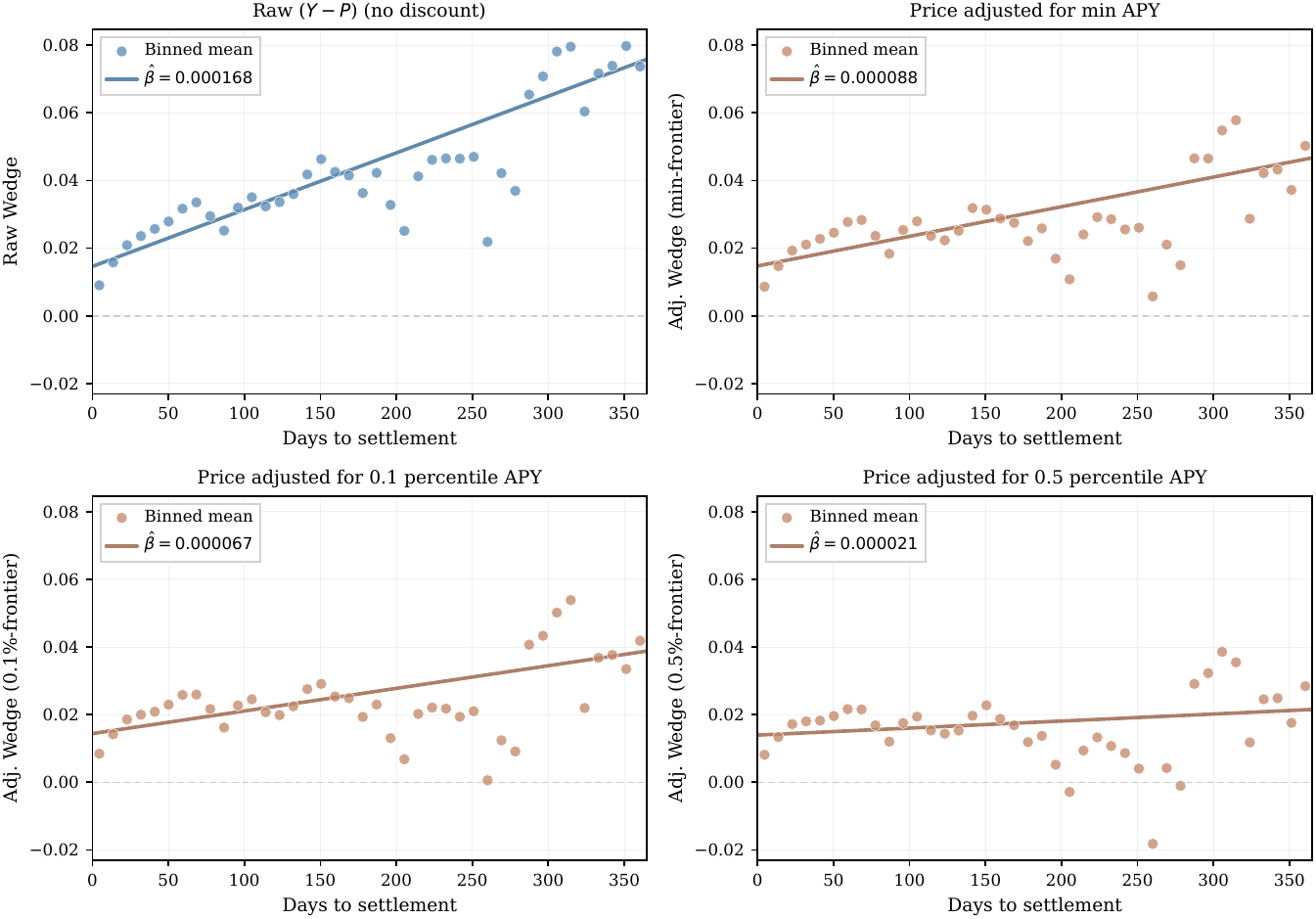}
    \caption{Binned visualization of the regression specifications corresponding to Table~\ref{tab:tail_regression_main}. The panels plot the raw wedge $X_i-P_{i,t}$ and the adjusted wedges using date-varying discount curves estimated from the minimum-ASW, 0.1-percentile, and 0.5-percentile near-certainty frontiers. Dots are binned means over horizon intervals up to 365 days, while red lines are OLS fits estimated on the full sample. Standard errors are clustered by market.}
    \label{fig:appendix-regression}
  \end{figure}

  Figure~\ref{fig:appendix-regression} visualizes the maturity-gradient
  attenuation underlying Table~\ref{tab:tail_regression_main}. The raw
  specification shows that longer-horizon near-certain contracts trade farther
  below eventual payoff, consistent with settlement-induced discounting
  increasing with horizon. The discount-adjusted specifications progressively
  flatten the gradient; attenuation increases monotonically as the lower-tail
  frontier is estimated more conservatively, with the 0.5-percentile frontier
  producing an adjusted slope statistically indistinguishable from zero.

  \subsection{Event-Level Frontier Portability}
  \label{app:crossfit-discount-robustness}

  A potential concern with the attenuation exercise is that the estimated
  discount curve $\hat D(\tau)$ is itself horizon-shaped and estimated from
  near-certain observations in the same structural environment. A curve
  estimated from that environment may mechanically attenuate the same horizon
  pattern it summarizes. We therefore repeat the main wedge regression using
  event-level cross-fitted discount curves: events are split into five folds,
  each frontier is estimated on the other four folds, and the resulting curve is
  applied only to held-out events before pooling the adjusted observations.

  \begin{table}[H]
    \centering
    \caption{All-event and held-out-event attenuation of the near-certainty
    horizon gradient. The raw slope is
    $\hat\beta_{\mathrm{raw}}=0.00016780$. Adjusted prices are capped at one
    after frontier adjustment. P-values are clustered by market.}
    \label{tab:crossfit-discount-robustness}
    \resizebox{0.85\linewidth}{!}{%
    \begin{tabular}{llrrr}
      \hline
      \textbf{Specification} & \textbf{Frontier} & \textbf{Adjusted slope} & \textbf{Slope reduction} & \textbf{p-value} \\
      \hline
      All events & Minimum & 0.00008757 & 47.8\% & 0.0079 \\
      All events & 0.1 percentile & 0.00006686 & 60.2\% & 0.0425 \\
      All events & 0.5 percentile & 0.00002076 & 87.6\% & 0.5285 \\
      \hline
      Held-out events & Minimum & 0.000073 & 56.4\% & 0.026 \\
      Held-out events & 0.1 percentile & 0.000055 & 67.2\% & 0.094 \\
      Held-out events & 0.5 percentile & 0.000012 & 92.9\% & 0.717 \\
      \hline
    \end{tabular}%
    }
  \end{table}

  The held-out-event estimates are the key portability test. Frontiers estimated
  from other event sets reduce the held-out horizon gradient by 56.4\% to
  92.9\%, slightly more than the corresponding all-event adjustments. Because
  attenuation does not weaken when the adjusted events are excluded from
  frontier estimation, the frontier is more consistent with a common
  platform-level settlement component than a purely same-sample statistical
  artifact.

  \subsection{Structural negRisk Payoff Identity}
  \label{app:neg-risk-identity}

  \paragraph{Claim (negRisk payoff identity).}
  Consider one event with $n$ mutually exclusive and exhaustive outcomes and
  realized outcome $\omega\in\{1,\dots,n\}$. Let
  \[
    Y_k(\omega)=\mathbf{1}\{\omega=k\},
    \qquad
    N_k(\omega)=1-\mathbf{1}\{\omega=k\}.
  \]
  For any subset $S\subseteq\{1,\dots,n\}$ with cardinality $m=|S|$, the
  NO-basket satisfies the exact payoff equivalence
  \begin{equation}
    \label{eq:neg-risk-identity}
    \sum_{k\in S}N_k\equiv (m-1)\cdot\mathbf{1}+\sum_{j\notin S}Y_j.
  \end{equation}
  Equation~\eqref{eq:neg-risk-identity} decomposes a NO basket into a
  deterministic component $(m-1)\mathbf{1}$ and a residual YES basket on the
  complementary outcomes. This identity underlies negRisk conversion: it allows
  an economically equivalent position to be transformed into cash-like
  collateral plus only the remaining complementary outcome exposure, compressing
  the capital that remains locked until settlement. In the common case
  $m=n-1$, the residual exposure is a single YES claim.

  \paragraph{Proof.}
  Define $\Pi_S:=\sum_{k\in S}N_k$. Its payoff is
  \[
    \Pi_S(\omega)
    =\sum_{k\in S}\left(1-\mathbf{1}\{\omega=k\}\right)
    =m-\mathbf{1}\{\omega\in S\}
    =(m-1)+\mathbf{1}\{\omega\notin S\}.
  \]
  Since outcomes are mutually exclusive and exhaustive,
  \[
    \mathbf{1}\{\omega\notin S\}=\sum_{j\notin S}Y_j(\omega).
  \]
  Substitution gives Equation~\eqref{eq:neg-risk-identity}. $\square$

  \subsection{Fee-Adjusted negRisk Conversion Bounds}
  \label{app:neg-risk-fees}

  Equation~\eqref{eq:neg-risk-identity} is an exact payoff identity. Conversion
  fees do not change that identity, but they reduce the net cash-like and
  residual YES outputs. The result is an idealized conversion-implied lower
  bound.

  \paragraph{Exact fee perturbation.}
  Let $\phi_c\in[0,1)$ denote the cash-leg fee rate and
  $\phi_y\in[0,1)$ the YES-leg fee rate. A conversion of the NO basket
  $\{N_k\}_{k\in S}$ with $|S|=m$ yields net outputs
  \begin{equation}
    \label{eq:fee-convert}
    \sum_{k\in S} N_k
    \longrightarrow
    (m-1)(1-\phi_c)\cdot \mathbf{1}
    +(1-\phi_y)\sum_{j\notin S} Y_j.
  \end{equation}
  In bips-based implementations, a single parameter maps to
  $\phi=\texttt{feeBips}/\texttt{FEE\_DENOMINATOR}$; when the split is unknown,
  $\phi_c=\phi_y=\phi$ is the reduced-form case.

  \paragraph{Reduced-form price bound.}
  Let $P(N_k,t)$ and $P(Y_j,t)$ denote time-$t$ prices and normalize USDC to
  one. Abstracting from execution frictions other than conversion fees, buying
  the NO basket and converting gives
  \begin{equation}
    \label{eq:fee-bound}
    \sum_{k\in S} P(N_k,t)
    \ge
    (m-1)(1-\phi_c)
    +(1-\phi_y)\sum_{j\notin S} P(Y_j,t).
  \end{equation}
  Under the reduced-form settlement-discount model
  $P(Y_j,t)=p_j(t)D(\tau)$, near-eliminated subsets with
  $p_S(t):=\sum_{k\in S}p_k(t)\approx 0$ satisfy
  \begin{equation}
    \label{eq:fee-bound-tail}
    \sum_{k\in S} P(N_k,t)
    \gtrsim
    (m-1)(1-\phi_c)+(1-\phi_y)D(\tau).
  \end{equation}

  \paragraph{Canonical $n$-market line.}
  In the common case $m=n-1$, let $\bar P_N(\tau;n)$ denote the average price
  of the $n-1$ NO legs in the basket. Equation~\eqref{eq:fee-bound-tail} gives
  the fee-adjusted conversion-implied line
  \begin{equation}
    \label{eq:nr-fee-price}
    \bar P_N(\tau;n)
    \gtrsim
    \frac{(n-2)(1-\phi_c)+(1-\phi_y)D(\tau)}{n-1}.
  \end{equation}
  Relative to the fee-free line $[(n-2)+D(\tau)]/(n-1)$, fees lower the price
  floor by $[(n-2)\phi_c+\phi_y D(\tau)]/(n-1)$ and therefore raise the
  annualized implied yield.

  \paragraph{Short-horizon implication.}
  A first-order expansion near par clarifies the maturity dependence of fees.
  With $D(\tau)\approx 1-r\tau$, the approximate continuously compounded
  implied rate is
  \begin{equation}
    \label{eq:nr-fee-rate-approx}
    r_{\mathrm{imp}}(\tau)
    \approx
    \frac{r}{n-1}+\frac{(n-2)\phi_c+\phi_y}{(n-1)\tau}.
  \end{equation}
  The first term is the compressed settlement-discount rate; the second is the
  fee term. Because the fee term scales as $1/\tau$, conversion fees matter
  disproportionately at short maturities, where small price wedges annualize
  sharply.

  \begin{figure}[H]
    \centering
    \includegraphics[width=\textwidth]{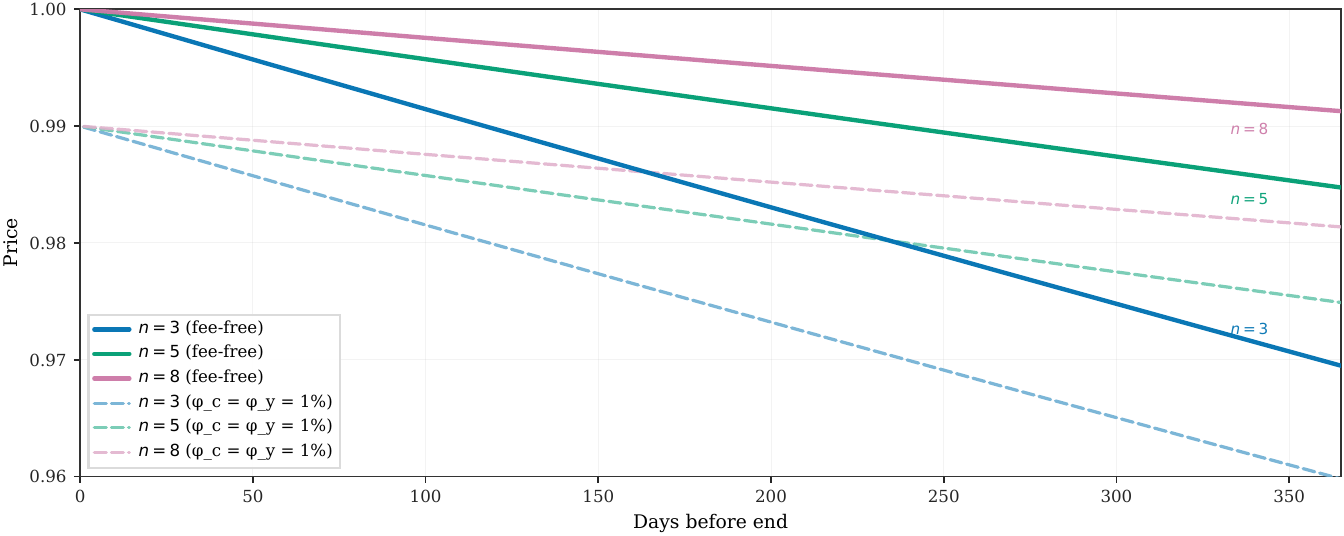}
    \caption{Fee-adjusted theoretical negRisk price floors. Solid colored curves show fee-free
    conversion-implied bounds for events with $n$ mutually exclusive outcomes;
    dashed colored curves set $\phi_c=\phi_y=1\%$. Larger $n$ and lower fees
    move the floor closer to par.}
    \label{fig:theoretical-price-fees}
  \end{figure}

  Figure~\ref{fig:theoretical-price-fees} isolates the fee channel in the
  negRisk mechanism. A NO basket converts into a cash-like component, subject
  to successful conversion and settlement, plus a residual YES claim. As $n$
  rises, the residual exposure becomes smaller relative to the cash-like part
  and the bound moves closer to par. Fees relax
  this idealized lower bound: the cash-leg fee loads on the deterministic
  component, while the YES-leg fee loads only on the discounted residual claim.
  In Polymarket's deployed NegRiskAdapter, a single market-specific
  \texttt{feeBips} parameter is applied to both outputs of conversion, making
  the symmetric case $\phi_c=\phi_y=\phi$ the closest reduced-form
  representation.\footnote{Polymarket NegRiskAdapter repository:
  \url{https://github.com/Polymarket/neg-risk-ctf-adapter}.}

  \subsection{Taker-Fee-Adjusted Execution Bounds}
  \label{app:neg-risk-taker-fees}

  The previous subsection studies fees charged inside the negRisk adapter. A
  separate friction arises before conversion: if the trader acquires the
  required legs with marketable orders, Polymarket taker fees raise the
  effective acquisition cost. This lowers the nominal posted-price
  bound at which the conversion trade breaks even. 

  For politics markets, let the taker-fee rate be \(f=0.04\)~\cite{Polymarket2026}. A marketable
  trade at price \(p\) pays per-share fee \(f p(1-p)\), so the effective
  acquisition price is
  \[
    p^{\mathrm{eff}}=p+f p(1-p).
  \]
  In a near-tail conversion, the same fee formula can be read as applying to a
  low-priced YES leg near \(p\approx0.01\), or equivalently to the
  complementary NO leg near \(1-p\approx0.99\). The traded-leg interpretation
  changes, but the fee schedule is symmetric in \(p\) and \(1-p\). The
  per-share taker fee is therefore approximately
  \[
    f p(1-p)\approx 0.04\cdot0.01\cdot0.99\approx0.000396.
  \]

  For a basket \(S\) with \(|S|=m\), the taker-executable conversion bound is
  \[
    \sum_{k\in S}
    \left[
      P(N_k,t)+fP(N_k,t)(1-P(N_k,t))
    \right]
    \ge
    (m-1)(1-\phi_c)
    +(1-\phi_y)\sum_{j\notin S}P(Y_j,t).
  \]
  Rearranging gives the nominal posted-price condition
  \[
    \sum_{k\in S}P(N_k,t)
    \ge
    (m-1)(1-\phi_c)
    +(1-\phi_y)\sum_{j\notin S}P(Y_j,t)
    -
    \sum_{k\in S}fP(N_k,t)(1-P(N_k,t)).
  \]

  In the canonical case \(m=n-1\), using the near-tail approximation
  \(P(N_k,t)\approx0.99\), the average nominal NO-price bound becomes
  \[
    \bar P_N^{\mathrm{taker}}(\tau;n)
    \gtrsim
    \bar P_N^{(\phi)}(\tau;n)-f\cdot0.99\cdot0.01,
  \]
  where
  \[
    \bar P_N^{(\phi)}(\tau;n)
    =
    \frac{(n-2)(1-\phi_c)+(1-\phi_y)D(\tau)}{n-1}.
  \]
  For politics markets,
  \[
    \bar P_N^{\mathrm{taker}}(\tau;n)
    \gtrsim
    \bar P_N^{(\phi)}(\tau;n)-0.000396.
  \]

  Taker fees therefore perturb the conversion-implied floor by about four basis
  points in price space near par. The level effect is small, but it is exactly
  the type of wedge that can become salient in short-horizon APYs because
  annualization scales a fixed price cost over a shrinking settlement horizon.

  \bibliographystyle{alpha}
  \bibliography{sample}
\end{document}